\documentclass{aa}
\usepackage[varg]{txfonts}
\usepackage{graphicx}
\usepackage{amsmath}
\usepackage{natbib}
\usepackage{hyperref}
\usepackage{xcolor}
\usepackage{wasysym}
\usepackage{gensymb}
\usepackage{caption}
\usepackage{subcaption}
\usepackage{float}
\usepackage{siunitx}

\newcommand{\HII}{\textrm{H~{\textsc{ii}}}}
\newcommand{\HI}{\textrm{H~{\textsc{i}}}}

\bibpunct{(}{)}{;}{a}{}{,}

\graphicspath{{image/}}

\begin{document}

\title{High angular resolution near-IR view of the Orion Bar revealed by Keck/NIRC2}

\author{Emilie Habart\inst{1}\and
  Romane Le Gal\inst{2,3}\and
  Carlos Alvarez\inst{4}\and
  Els Peeters\inst{5,6,7}\and
  Olivier Bern\'e\inst{8}\and
  Mark G. Wolfire\inst{9}\and
  Javier R. Goicoechea\inst{10}\and
  Thi\'ebaut Schirmer\inst{11}\and
  Emeric Bron\inst{12} \and
Markus R\"ollig \inst{13}
}

\institute{Universit\'e Paris-Saclay, CNRS, Institut d'Astrophysique Spatiale, 91405 Orsay, France\and
Institut de Plan\'etologie et d'Astrophysique de Grenoble (IPAG), Universit\'e Grenoble Alpes, CNRS, F-38000 Grenoble, France\and
Institut de Radioastronomie Millim\'etrique (IRAM), 300 rue de la piscine, F-38406 Saint-Martin d'H\`{e}res, France\and
W.M. Keck Observatory, 65-1120 Mamalahoa Hwy, Kamuela, HI, USA\and
Department of Physics \& Astronomy, The University of Western Ontario, London ON N6A 3K7, Canada\and
Institute for Earth and Space Exploration, The University of Western Ontario, London ON N6A 3K7, Canada\and
Carl Sagan Center, SETI Institute, 339 Bernardo Avenue, Suite 200, Mountain View, CA 94043, USA\and
Institut de Recherche en Astrophysique et Plan\'etologie, Universit\'e de Toulouse, CNRS, CNES, UPS, 9 Av. du colonel Roche, 31028 Toulouse Cedex 04, France\and
Astronomy Department, University of Maryland, College Park, MD 20742, USA\and
Instituto de F\'{\i}sica Fundamental (IFF), CSIC,  Calle Serrano 121-123, 28006, Madrid, Spain\and
Department of Space, Earth and Environment, Chalmers University of Technology, Onsala Space Observatory, 439 92 Onsala, Sweden\and
LERMA, Observatoire de Paris, PSL Research University, CNRS, Sorbonne Universit\'es, F-92190 Meudon, France\and
I. Physikalisches Institut der Universit\"{a}t zu K\"{o}ln, Z\"{u}lpicher Stra{\ss}e 77, 50937 K\"{o}ln, Germany
}

\date{2022}

\abstract
% context heading (optional)
  % {} leave it empty if necessary  
  {Nearby Photo-Dissociation Regions (PDRs), where the gas and dust are heated by the far UV-irradiation emitted from stars, are ideal templates to study the main stellar feedback processes. 
  }
  % aims heading (mandatory)
   {With this study we aim to probe the detailed structures at the interfaces between ionized, atomic, and molecular gas in the Orion Bar. This nearby prototypical strongly irradiated PDR will be among the first targets of the James Webb Space Telescope (JWST) within the framework of the PDRs4All Early Release Science program.
   }
  % methods heading (mandatory)
   {We employed the sub-arcsec resolution accessible with Keck-II NIRC2 and its adaptive optics system to obtain the most detailed and complete images, ever performed, of the vibrationally excited line H$_2$ 1-0 S(1) at 2.12~$\mu$m, tracing the dissociation front, and the [FeII] and Br$\gamma$ lines, at 1.64 and 2.16~$\mu$m respectively, tracing the ionization front. The former is a powerful tracer of the FUV radiation field strength and gas density distribution at the PDR edge, while the last two trace the temperature and density distribution from the ionized gas to the PDR. We obtained narrow-band filter images in these key gas line diagnostic over $\sim 40''$ at spatial scales of $\sim$0.1$''$ ($\sim$0.0002~pc or $\sim$40~AU at 414~pc).}
  % results heading (mandatory)
   {The Keck/NIRC2 observations spatially resolve a plethora of irradiated sub-structures such as ridges, filaments, globules and proplyds. 
   This portends what JWST should accomplish and how it will complement the highest-resolution ALMA maps of the molecular cloud. A remarkable spatial coincidence between the H$_2$ 1-0 S(1) vibrational and HCO$^+$ J=4-3 rotational emission previously obtained with ALMA is observed. This likely indicates the intimate link between these two molecular species and highlights that in high pressure PDR the H/H$_2$ and C$^+$/C/CO transitions zones come  closer as compared to a typical layered structure of a constant density PDR.
   The H/H$_2$ dissociation front appears as a highly structured dense region with over-dense sub-structures with a typical thickness of a few $\sim$10$^{-3}$~pc. 
   This is in agreement with several previous studies  that claimed that the Orion Bar edge is composed of very small, dense, highly irradiated PDRs at high thermal pressure immersed in a more diffuse  environment.}
  % conclusions heading (optional)
   {}

\maketitle

\section{Introduction}

Stellar feedback influences the star formation activity through molecular cloud dissolution, compression processes and increase in ionization fraction, which slows down ambipolar diffusion.
However, the precise nature and importance of all theses mechanisms relative to other processes that influence star formation remains largely unknown.
Stellar feedback can take the form of radiative heating and pressure, thermal expansion and evaporation, and stellar winds and shocks \citep[e.g.,][and reference therein]{Pabst20,Schneider2020}. The combined radiative and dynamical feedback of massive stars on their parental cloud is not yet understood.
Therefore, direct observation and quantification of the radiative heating efficiency (e.g., the coupling of gas and dust to the intense stellar-ultraviolet fields) and the mechanical energy injection is necessary.

%Such observations are ideal for tracing outflows and shock waves in star-forming regions where sub-arcsecond resolution uncovers morpologies crucial to interpreting the physics at work in these systems \citep[e.g.][]{Bally2015}.
Recent near-infrared images obtained with the Gemini telescope at high angular resolution ($\sim 0.06-0.11''$) revealed, at a spectacular level of detail, unexpected structures within the strongly irradiated molecular cloud at the Western Wall in  Carina at a distance of 2.3~kpc \citep{Hartigan2020}. These authors found series of ridges, fragments, and waves with sizes in the $\sim$200-2000 AU range that may result from the complex dynamical and radiative processes that sculpt these systems. Indeed, while regularly-spaced ridges that run parallel to the photodissociation front may suggest large-scale magnetic fields are dynamically-important \citep{Mackey2011}, several fragments %of width $\sim$ 1800 au 
and waves 
%with a half-wavelength of 2500 au 
could resemble Kelvin-Helmholtz instabilities. 
\cite{Berne2010} reported the presence of a series of five surprisingly regular wavelets (separated by 0.1 pc or 20000 AU) at the surface of the Orion molecular cloud. These authors proposed that these waves formed by the mechanical interaction of high-velocity plasma/gas, produced by massive stars, with the dense molecular gas, which has provoked hydrodynamical instabilities. 
%suggested hydrodynamical response of gas in the ISM to strong stellar action.
%From Spitzer observation, \cite{Berne2010} also reported the presence of a series of five surprisingly regular wavelets (separated by 0.1 pc or 20000 au) at the surface of the Orion molecular cloud near where massive stars are forming. They proposed that these waves have been formed by the mechanical interaction of high-velocity plasma/gas, produced by massive stars, with the dense molecular gas, which has provoked hydrodynamical instabilities. 
%The waves seem to be a Kelvin-Helmholtz instability that arises during the expansion of the nebula as gas heated and ionized by massive stars is blown over pre-existing molecular gas. 
Sub-arcsecond resolution is required to resolve detailed structures even for the nearest regions of massive star formation. This capability now exists with adaptive optics (AO) imaging, and will be further developed/expanded in the near future with JWST.

The main goal of the present study is to observe and quantify the stellar feedback in the closest site of ongoing massive star-formation: the Orion molecular cloud. This region is located about 5 times closer than the Carina star forming region. 
The dominant stellar feedback processes can be probed by observations of Photo-Dissociation Regions (PDRs) where far-ultraviolet photons of massive stars create warm regions of gas and dust in the neutral atomic and molecular gas \citep[e.g., for a recent review see][and references therein]{wolfire22}. Nearby, edge-on PDRs such as the Orion Bar are ideal targets due to their wide extension on the sky, bright emission, and proximity that allow to probe these processes on a small physical scale. % providing exquisite high-sensitivity and spatially resolved observations. 
PDRs are particularly bright in the infrared (IR). 

However, until now a very high angular (sub-arcsec resolution) IR molecular view of the Orion Bar PDR was missing. 
In this study, we present observations obtained with Keck/NIRC2 using AO, which provide the most complete and detailed maps of the complex UV-irradiated region where the conversion from ionized to atomic to molecular gas occurs.
We mapped the vibrationally excited line of H$_2$ at 2.12 $\mu$m, tracing the dissociation front, and the [FeII] line at 1.64 $\mu$m and the Br$\gamma$ line at 2.16 $\mu$m, tracing the ionization front.
The proximity of the region\footnote{The most commonly adopted distance to the Bar is 414\,pc \citep{Menten07} although more recent observations, including Gaia, point to slightly lower
values \citep{Kounkel17,Gross18}.} (414~pc) combined with recent advances in instrumentation allows for the first time to conduct studies from the (sub-)parsec scales to the smallest structures of $\sim$100 AU. Observations with AO provide nearly diffraction-limited images in the near-IR over a field of view of $1'$ in size. 
%In this paper, we present observations obtained with the Keck/NIRC2 telescope of the well-studied Photo-Dissociation Region (PDR) the Orion Bar, located about 5 times closer than the Carina star forming region. 

% The Keck/NIRC2 observations improve the spatial resolution of the available H$_2$ \citep[e.g.,][]{vanderWerf96,Walmsley00}
% and ALMA molecular emission maps ($\sim1''$) \citep{Goico16} by a factor of ten.
%Furthermore, the Keck/NIRC2 observations presented in this paper were used to optimize an accepted JWST Early  Release Science (ERS) proposal on the Orion Bar PDR \citep{paper_ERS_PASP_2022}.
% %\footnote{\url{https://www.stsci.edu/jwst/science-execution/approved-programs/dd-ers/program-1288
% %https://jwst.stsci.edu/observing-programs/approved-ers-programs/program-1288}}. \footnote{\url{https://www.stsci.edu/jwst/science-execution/approved-programs/dd-ers/program-1288}}.
% %one of the 13 accepted ERS programs called ``PDRs4all: Radiative feedback from massive stars"  (ID1288) which is dedicated to the study of radiative feedback from massive stars.
% It helped us to accurately determine the line intensity peak in order to adjust the integration time, get a high S/N and avoid saturation problems. 
% %In this ERS program, NIRCAM imaging in several narrow gas lines filters will be obtained over a much larger field of view, as well as, IFU NIRSpec and MIRI mosaic towards the Bar.
% %Keck give an idea of which Small scale structures will be probed by JWST / how it complement ALMA data.

The paper is organized as follows. 
%We discuss the target in Section~\ref{sec:target}. 
The target the Orion Bar is described in Sect.~\ref{sec:orion-bar}.
The observations and data reduction are described in Sect.~\ref{sec:obs}. In Sect.~\ref{sec:spatial-distribution}, the spatial morphology of the line intensities is discussed and compared to previous observations. In Sects.\ \ref{sec:intensity-extinction} and ~\ref{sec:model-predictions}, the observed line intensities are corrected for dust extinction and compared to the model predictions. Finally, a short summary and prospects for JWST are given in Section~\ref{sec:conclusions}. 

\section{The Orion Bar}
\label{sec:orion-bar}

The Orion Bar, a strongly ultraviolet (UV) irradiated PDR,  is an escarpment of the Orion molecular cloud (OMC), the closest site of ongoing massive star formation.  
%The gas density in the ambient molecular cloud is estimated to be $n_H = (0.5-1.0~10^5$ cm$^{-3}$ \citep{Tielens_1985b,Hoger95}. 
The Bar is illuminated by the \mbox{O7-type} star \mbox{$\theta^1$ Ori C}, the most massive and luminous member of the Trapezium  cluster at the heart of the Orion Nebula \citep[e.g.,][]{Odell01}. The trapezium cluster creates a blister \HII\, region that is eating its way into the parental cloud.
 A large cavity has been carved out of the molecular gas and the inner concave structure tilts to form the Orion Bar \citep{Wen1995, Odell01}. 
%Then the main ionization front flattens, but not so much that it is still illuminated directly by the ionizing star.
\cite{weilbacher2015} presented MUSE/VLT integral  spectroscopic data of the central part of the Orion Nebula %over $5' \times 6'$ 
 with an angular resolution of $\sim$0.7 to 1.2$''$. %and wavelength range from 4595 to 9366 Angstrom. 
They derived an extinction map and estimated physical properties (electron temperature and density) of the Orion Nebula HII region. 
The Far-UV (FUV) radiation field incident on the Orion Bar PDR is G$_0= 1-4\times10^4$ in Habing units \citep[e.g.,][]{Marconi98}. 
Beyond the ionization front (IF), where the gas converts from fully ionized to fully neutral, only FUV photons with energies below 13.6\,eV penetrate the cloud. This corresponds to the edge of the PDR. 
Many previous works have studied the Orion Bar PDR's spatial stratification. 
The first PDR layers are predominantly neutral and atomic:  \mbox{[H]\,$>$\,[H$_2$]\,$\gg$\,[H$^+$]}. 
They  display NIR atomic emission lines from low ionization potential elements \citep[][]{Walmsley00} as well as carbon and sulfur radio recombination lines \citep[][]{Wyrowski97,Cuadrado19,Goicoechea21}.  
This warm and moderately dense gas ($n_{\rm H}$ of a few 10$^4$\,cm$^{-3}$)  is  mainly cooled by the very bright FIR [C$^+$]\,158\,$\mu$m and [O$^0$] 63 and 145\,$\mu$m fine-structure lines \citep[][]{Tielens93,Herrmann97,Bernard-Salas12,Ossenkopf13}.
The atomic PDR zone also  hosts the peak of the Mid-IR aromatic particles emission \citep[e.g.,][]{Bregman89,Sellgren90,Tielens93,Giard94,Knight21,Schirmer2022}.  

At about 10-15$''$ (or 0.02-0.03~pc) from the IF (at \mbox{$A_V$\,$\simeq$\,0.5-2~mag} of visual extinction into the neutral cloud), the dissociating FUV photons are sufficiently attenuated and most of the hydrogen becomes molecular\footnote{This range of extinction value is only value for high $G_0$/n$_H$ PDRs like the Orion Bar, but not in lower $G_0$/n$_H$ PDRs where the H/H$_2$ transitions occur at much lower $A_v$ due to the H$_2$ self-shielding.} \citep[e.g.,][]{vanderWerf13}. 
This corresponds to the dissociation front (DF) where the H/H$_2$ transition takes place. The DF displays a forest of IR rotational and vibrationally excited H$_2$ lines \citep[e.g.,][]{Parmar91,Luhman94,vanderWerf96,Allers05,Shaw09,Zhang21}, including \mbox{FUV-pumped} levels up to $v$\,=\,12 \citep{Kaplan17, Kaplan2021} and HD rotational lines \citep{Wright99,Joblin18}. 
FUV-pumped H$_2$ drives the formation of molecular ions such as CH$^+$ and SH$^+$ through reactions of C$^+$ and S$^+$ with vibrationally excited \citep[e.g.,][and references therein]{Goicoechea17,Parikka17,Lehmann22}. %These molecules are also routinely(?) detected in distant galaxies by ALMA and Herschel {\bf[ref. needed here]}.  
Analysis of the IR H$_2$ and 21\,cm \HI~lines suggests warm temperatures (\mbox{$T_{\rm k}$\,$\simeq$\,400-700\,K}) at the DF. 
Using stationary PDR models (i.e. PDR models with stationary chemistry and without dynamics), the gas density in the atomic gas of $n\rm{_H} = 4-5~10^4$ cm$^{-3}$ is consistent with the [OI] and [CII] line emission and the separation between the IF and DF \citep{Tielens93,Hoger95,Marconi98}.

The IF and DF separation depends also on the dust extinction. 
Larger grains than found in diffuse ISM that have lower FUV absorption cross-sections need to be invoked in isobaric PDR models in order to be consistent with the observed IF and DF separation of $\sim$10-15$''$ \citep[e.g.,][]{Allers05}.  Reduction in the FUV attenuation cross-section is in agreement with the recent work of \cite{Schirmer2022} modeling dust emission and extinction throughout the Orion Bar PDR. Analyzing the Spitzer and Herschel observations with the radiative transfer code SOC \citep{Juvela2019}, together with the dust THEMIS model \citep{Jones2013,jones2017}, \cite{Schirmer2022} found that nano-grains are in fact strongly depleted compared to diffuse ISM (by a factor $\ge$10), leading to a FUV attenuation reduction, and a minimum grain size that is larger than in the diffuse ISM. In their model, the gas density is estimated to be about 4~10$^4$ cm$^{-3}$ at the mid-IR aromatic nano-dust emission peak, and about 1.5~10$^5$ cm$^{-3}$ on average between the mid-IR emission peak and the H$_2$ emission peak. This is consistent with previous estimates. 
%FUV extinction dust properties (which  vary as function of cloud depth).

Beyond the DF, between \mbox{$A_V$\,=\,2} and 4\,mag, the C$^+$/C/CO transition takes place  \citep[][]{Tauber95} and the PDR becomes molecular. Some observations and models suggest the presence of relatively large (0.01-0.02 pc) high density clumps (\mbox{$n_{\rm H}$\,=\,10$^6$--10$^7$cm$^{-3}$}) in the molecular PDR \citep[e.g.,][]{Burton90,vanderWerf96,YoungOwl00,Lis03,Andree17}. 
These over-dense components must be embedded in a lower density medium mainly responsible for the extended PDR emission. 
Alternatively, the observed H$_2$ and high-J CO lines emission close to the DF 
may be explained by a roughly isobaric PDR, at high thermal-pressure \mbox{$P_{\rm th}/k$\,$\approx$\,10$^8$\,cm$^{-3}$\,K} \citep[e.g.,][]{Allers05,Joblin18}.
%, in which the gas density naturally rises as the gas cools
%in which the contrasting densities that were derived from different tracer are naturally explained by the strong density gradient in the PDR that is induced by the temperature gradient.
We note that high density and pressure appear as a
common feature in high FUV field PDRs, as demonstrated by a number of observations \citep[e.g.,][]{Ossenkopf2010,Sheffer11,Kohler2015,Wu2018}. 
%H2 line emission (Sheffer et al. 2011), and [OI] and high-J CO molecular line emission (Ossenkopf et al. 2010, Kohler et al.  Wu et al. 2018).
 In the high pressure isobaric model, a constant pressure from the atomic to the warm molecular region results in an atomic region and a very narrow emission for the warm molecular tracers. In the \cite{Joblin18} model with P=2.8 10$^8$\,cm$^{-3}$\,K the atomic region had a size of $\sim$3.5$''$ while the emission zone of the H$_2$ (1-0) S(1) line has a width of $\sim$0.25$''$. The predicted atomic region width is nevertheless  significantly smaller than the one observed ($\sim$10-15$''$) and the very narrow molecular emission is unresolved in most tracers until now.
%However, the high pressure layers of a PDR are predicted to be very narrow ( $\sim 1-2''$) and thus unresolved in most tracers until now.
%Several studies as \citep{Joblin18,Parikka17} claimed that the Orion Bar is composed by very small ($\sim$1-2'') dense highly irradiated PDRs with differing local density immersed in a less dense environment spatially extended ($\sim$10-20ÕÕ).

 While most of the previous molecular studies relied on modest angular resolution observations, ALMA has provided \mbox{$\sim$\,1\,$''$} resolution images of the molecular emission \citep{Goico16}. Instead of an homogeneous PDR with well-defined and spatially separated \mbox{H/H$_2$} and \mbox{C$^+$/C/CO} transition zones, ALMA revealed rich small-scale ($\sim$0.004 pc) over-dense structures (akin to filaments and globules), sharp edges, and bright emission from an embedded proplyd \citep[object \mbox{203-506};][]{Champion17}.
 These observations have challenged the traditional view of PDRs \citep[and their models, e.g.,][]{Kirsanova19} and revealed a steeply varying interface between the atomic and the molecular gas \citep[Fig. 1c,][]{Goico16}. The observed over-dense substructures may have been induced by UV radiation-driven compression \citep{Gorti02,Tremblin2012}.
Advection of the molecular gas through the DF has been suggested by CO emission observed by ALMA in the form of globules or plumes apparently extending into the atomic gas \citep{Goico16}. 

Recent FIR observations of the dust emission polarization from the Orion Bar PDR suggest a relatively modest plane-of-the-sky magnetic field strength of $B_0$\,$\simeq$\,300\,$\mu$G \citep[e.g.,][]{Chuss19,Guerra21}.
Hence, magnetic pressure of 3~10$^7$ K cm$^{-3}$ (comparable to the ambient and inter-clump thermal pressure of 1-3 ? 10$^7$ K cm$^{-3}$) may also play a role in the PDR dynamics \citep[e.g.,][]{Pellegrini09,Pabst20}.

\begin{figure}%[h]
    %\centering
        \resizebox{\hsize}{!}{\includegraphics{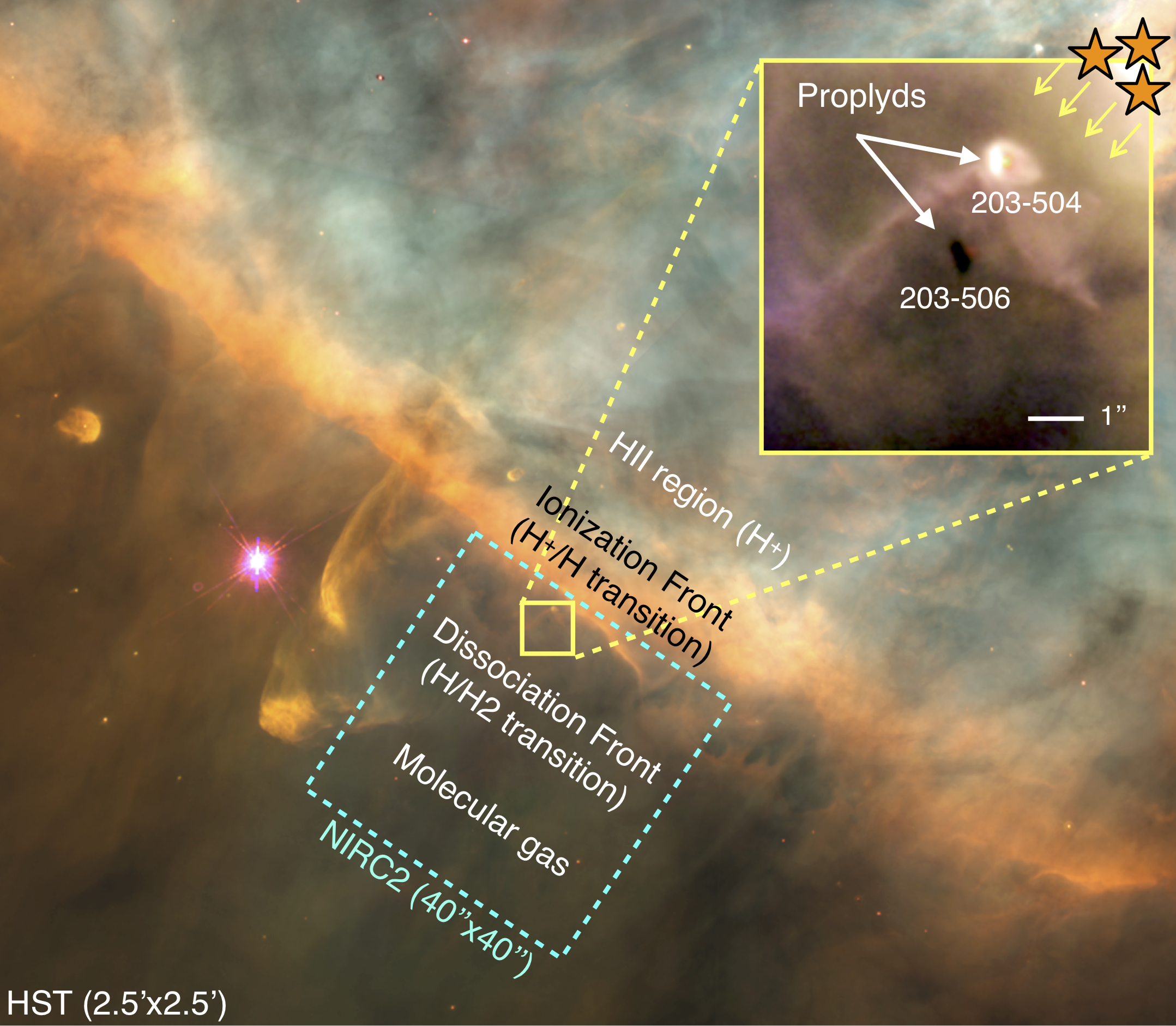}}
%    \resizebox{\hsize}{!}{\includegraphics{images/Figure1-Keckpaper.png}}
    %\includegraphics[width=0.65\textwidth]{images/Figure1-Keck-paper.png}
    \caption{Hubble Space Telescope WFPC2 mosaic of the Orion Bar %combining 45 images obtained in blue, red, and green filters 
    (Credit: NASA/STScI/Rice Univ./C.O'Dell et al. - Program ID: PRC95-45a). The NIRC2 wide camera FoV is shown with the cyan square. The yellow square zooms into the two proplyds lying in the targeted FoV.}
    \label{fig:HST}
\end{figure}

\begin{figure}[h]
    \centering
     \includegraphics[width=0.5\textwidth]{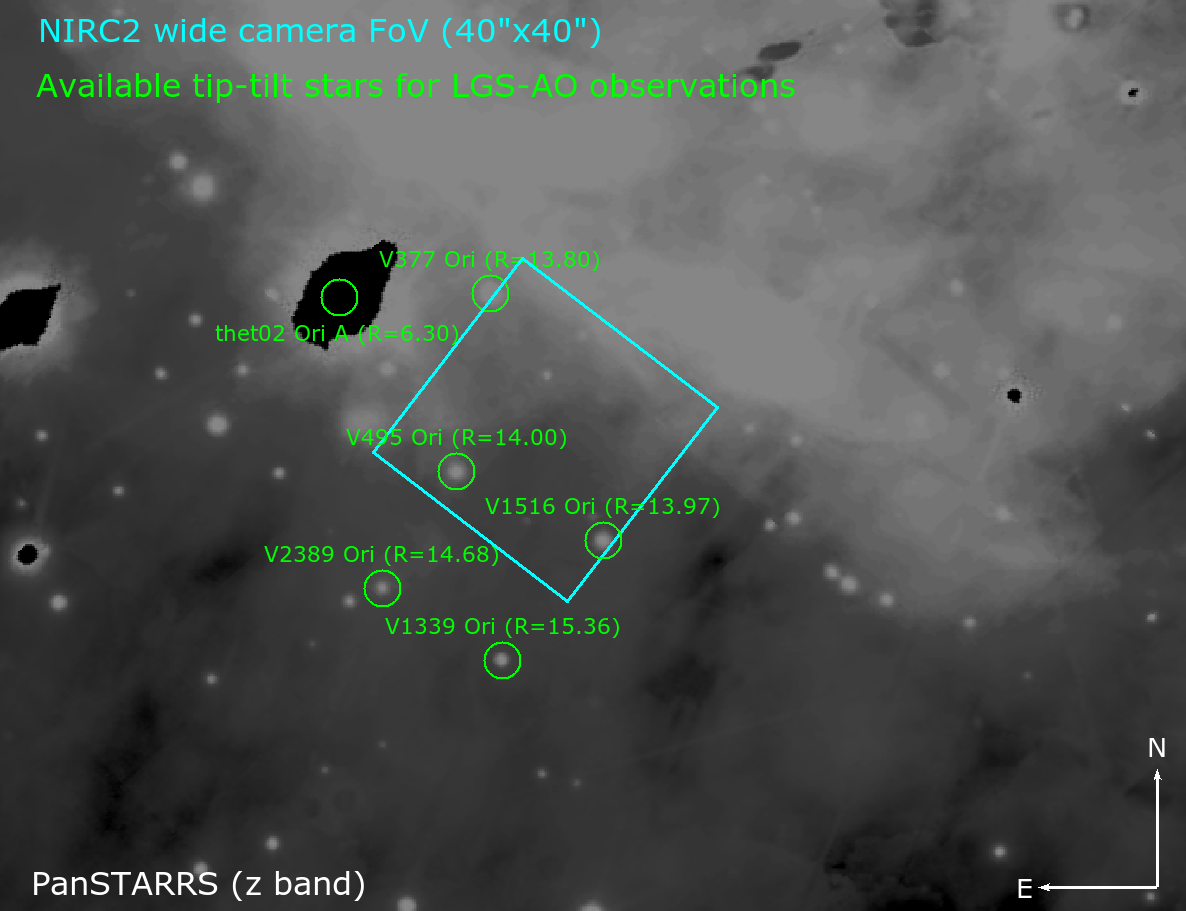}
    \caption{The PanSTARRS (z band) image showing the NIRC2 wide camera FoV (cyan square, $40''\times40''$) centered on RA=$05:35:20.662$, DEC=$-05:25:19.38$ (J2000). 
     The green circles enclose tip-tilt stars that can be used to perform the LGS observations. 
        }
    \label{fig:Keck-FOV}
\end{figure}

\section{Observations and data reduction}
\label{sec:obs}

\subsection{Observations}

On December 4, 2020 (Program ID N004), we used the NIRC2 wide field camera in combination with the Keck-II AO Laser Guide Star (LGS) system to observe a representative Field-of-View (FoV) of $40''\times40''$ of the Orion Bar centered on RA=05:35:20.662, DEC=-05:25:19.38 (J2000) with a pixel size of $0.04''$, as illustrated in Figs.~\ref{fig:HST} and ~\ref{fig:Keck-FOV}. This instrumental setup provides the optimal combination of FoV and diffraction-limited imaging capabilities to fulfill our science goals. %of our program. 
The FoV was rotated by 52.5$^{\degree}$ (North to East). Observations were obtained in the narrow-band filters H$_2$ $v=1-0$ ($\lambda_{c}=2.128\,\mu$m, $\Delta \lambda=0.034\,\mu$m), [FeII] ($\lambda_{c}=1.645\,\mu$m, $\Delta \lambda=0.026\,\mu$m) and Br~$\gamma$ ($\lambda_{c}=2.168\,\mu$m, $\Delta \lambda=0.0326\,\mu$m).
We also obtained observations in the Kcont ($\lambda_{c}=2.271\,\mu$m, $\Delta\,\lambda=0.033\, \mu$m) and Hcont ($\lambda_{c}=1.580\,\mu$m, $\Delta \lambda=0.023\,\mu$m) filters to subtract the continuum from the H$_2$ $v=1-0$, Br~$\gamma$, and [FeII] line maps respectively. 

Fig.~\ref{fig:Keck-FOV} shows the different point sources within 60$''$ of the center of the region that can be used as tip-tilt stars for the LGS system.
The executed observation sequence was a 5-point dither pattern using a $\sim1''$ dither step with H$_2$ $v=1-0$ and Kcont consecutive images on each dither position. The same sequence was executed for the [FeII] and Hcont filters, and the Br~$\gamma$ and Kcont filters.  
The exposure time per image was 150s (15s $\times$ 10 coadds) in multiple-correlated double-sampling mode with 8 reads per coadd.

We integrated 750s on-source for each filter. The observing strategy was successful although it was a bit challenging due to the variable nebular background on the tip-tilt star. In fact, the nebular background can significantly increase the effective magnitude of the tip-tilt star, which in turn can degrade the AO correction. In an environment with such a complex background as the Orion Bar, the automatic AO acquisition sequence may need to be modified to manually take a background for the tip-tilt sensor away from the nebular region. Such manual background acquisition reduces the efficiency of the AO observations. We took images in the H$_2$ $v=1-0$ and Kcont filters with V377 Ori as the tip-tilt star. However, due to the high nebular background, we used $\theta ^2$ Ori A as the tip-tilt star for the [FeII]/Hcont and Br-gamma/Kcont sequences. The factor of $\sim$2 larger separation between $\theta ^2$ Ori A and the science FoV with respect to V377 Ori was compensated by the fact that $\theta ^2$ Ori A is 7.5~magnitudes brighter on the wave front sensor.
Our observations clearly benefited from good seeing conditions and AO correction (K-band Strehl ratio $\sim$ 30\%). 
%Strelh ratio = 30\%
The full width at half maximum (FWHM) of the point-spread function (PSF) was $\sim 0.11''$. For comparison, JWST will achieve a resolution of 0.085'' in the K-band. 
In summary, these unusual bright extended source observations for NIRC2 were successful. 

\subsection{Data reduction}

The basic data reduction of the imaging frames (i.e., dark correction, flat field, sky substraction) were performed using standard procedures for near-IR imaging with a modified version of the reduction pipeline from the UCLA Galactic Center group based on Pyraf \citep{Lu2009}. 
The set of frames slightly offset in telescope position were realigned and co-added to produce the final image. 
We calculated star centroids and offsets from one frame to the other using the IRAF task IMALIGN in order to line up the images. Thus, the images were co-added by weighted  median which suppresses the ghosts and the diffraction spike due to the bright $\theta ^2$Ori A star. 
 The final images are of dimension of $38.64''\times 38.88''$ or 0.08~pc $ \times 0.08$~pc at 414~pc. 

In order to overcome the telescope pointing error (which is of the order of $0.5''$) and determine the proper World Coordinate System (WCS), we matched the stars in our NIRC2 image to Gaia coordinates. We used the Gaia catalog data early release 3 (https://www.cosmos.esa.int/web/gaia).

For the flux calibration, we determined the conversion factor between the detector counts and physical units using previous observations of the H$_2$ 1-0 S(1) line obtained with NTT/SOFI \citep{Walmsley00}.
We convolved the continuum-subtracted Keck/NIRC2 H$_2$ line emission to the beam of the NTT/SOFI observation (seeing limited $\sim 1''$). We estimated the flux conversion factor in regions of the maps with high signal to noise ratio. 
Thus, the reduced science maps (in Analog Digital Unit per second) were multiplied by this conversion factor to get the flux-calibrated science maps in erg cm$^{-2}$ s$^{-1}$ sr$^{-1}$. We assigned an accuracy of about 25\% to the measurements of the line intensity.
The same conversion factor was used for the H$_2$ $v=1-0$, Br~$\gamma$ and [FeII] lines maps since they were obtained with almost the same weather conditions and the detector response is similar. We compared the fluxes of each of the lines we obtained with those measured by \cite{Walmsley00} through the PDR (see their position C, Table 1) and found a good agreement considering the different spatial resolution. 
%For the H$_2$ 1-0 S(1) line, we also compared our calibrated map with the BEAR observations and found an agreement within \textbf{??\%}. 
Maps that are not continuum-subtracted were divided by the filter width and are in units
erg cm$^{-2}$ s$^{-1}$ sr$^{-1}$ $\mu$m$ ^{-1}$. 
We must underline that we could have fluxed-calibrated the observations using the photometric standard FS-7 observed in  the H and Ks filters. However, we would obtained similar uncertainties due to the sky background variations between the standard star observations and the science observations, which was significant in the K band.

\section{Spatial distribution}
\label{sec:spatial-distribution}

In this section, the spatial morphology of the hydrogen Brackett $\gamma$, iron and molecular hydrogen lines are presented and compared to previous observations.

\begin{figure*}[h!]
    \centering
   % \subfloat[][]{
  % \rule{0.45\linewidth}{8cm}} 
  %\subfloat[][]{ \rule{0.45\linewidth}{8cm}} 
       \includegraphics[width=0.45\textwidth]{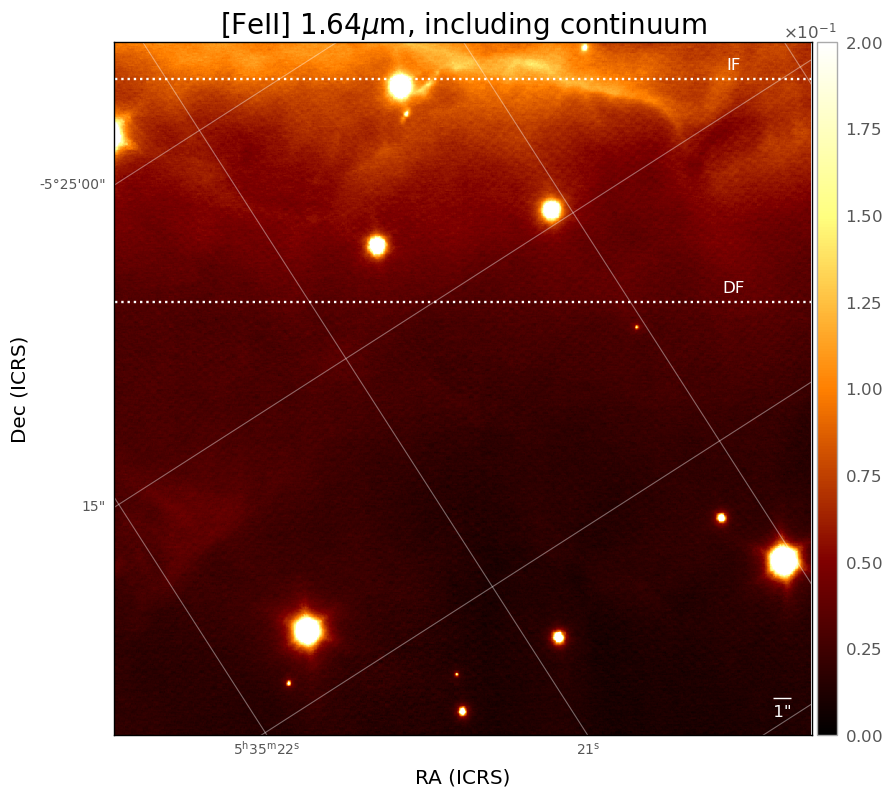}
              \includegraphics[width=0.45\textwidth]{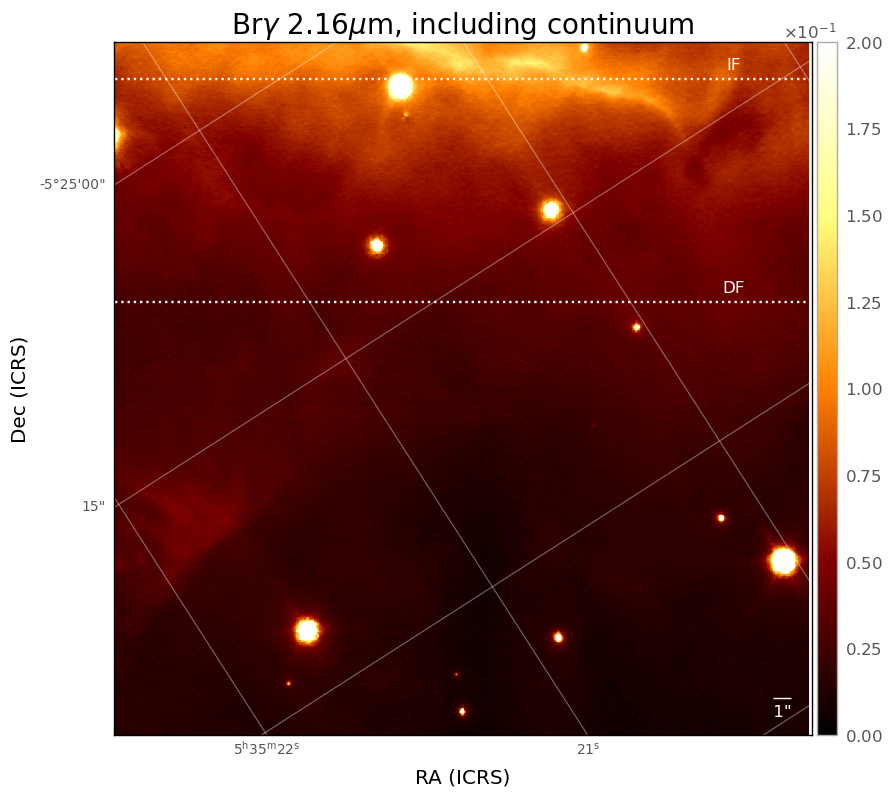}
                     \includegraphics[width=0.45\textwidth]{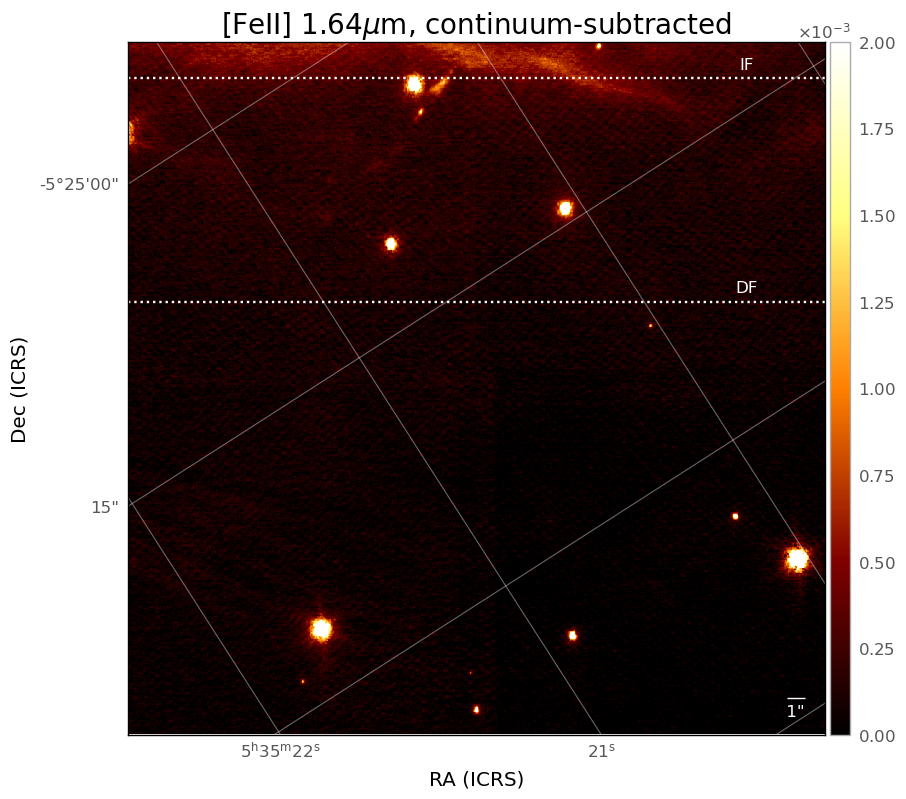}
                            \includegraphics[width=0.45\textwidth]{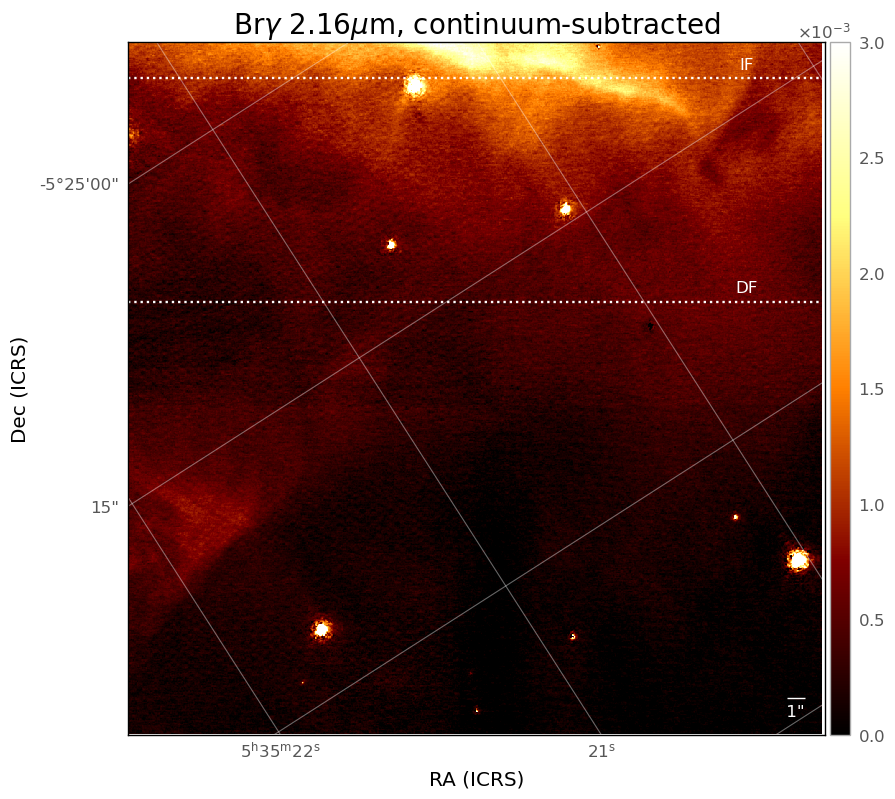}
      \caption{Integrated intensity maps of the [FeII] and continuum-subtracted [FeII] line emission (left column) and of the Br$\gamma$ and continuum-subtracted Br$\gamma$ line emission (right column). Units are erg cm$^{-2}$ s$^{-1}$ sr$^{-1}$ $\mu {\rm m}^{-1}$ for the [FeII] and Br$\gamma$ maps and erg cm$^{-2}$ s$^{-1}$ sr$^{-1}$ for the continuum-subtracted line emission map. The horizontal dashed lines indicate the average position of the ionization and dissociation fronts (as determined from the averaged emission cut over the entire FoV presented in Fig.~\ref{fig:Keck-cut}).}
       \label{fig:Keck-map}
\end{figure*}

      \begin{figure*}[h!]
    \centering
          \includegraphics[width=0.65\textwidth]{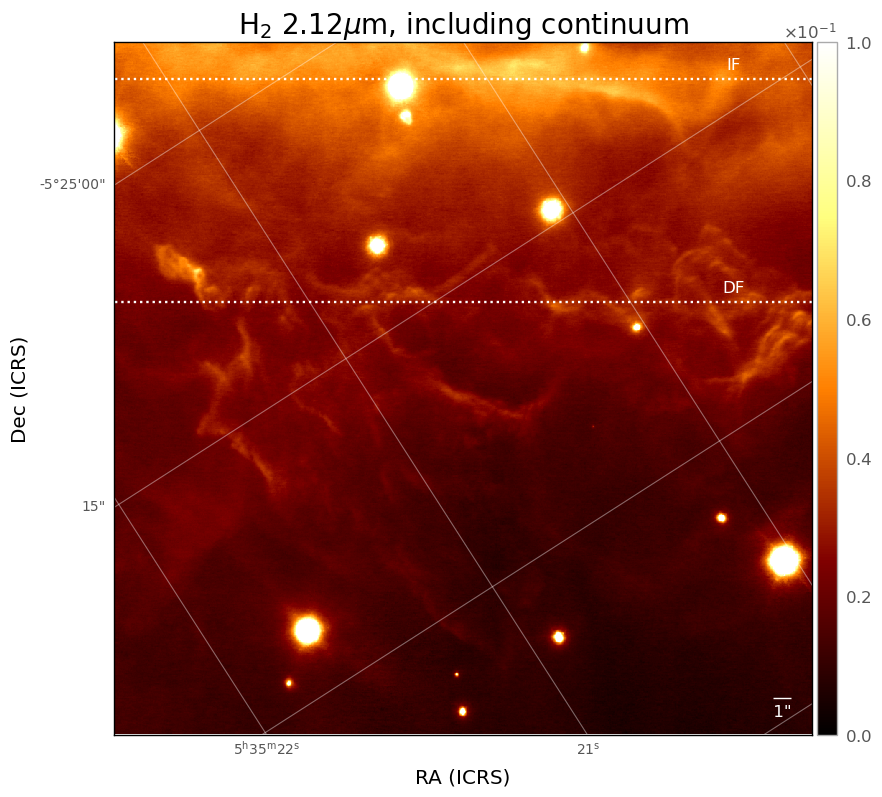}
                   \includegraphics[width=0.65\textwidth]{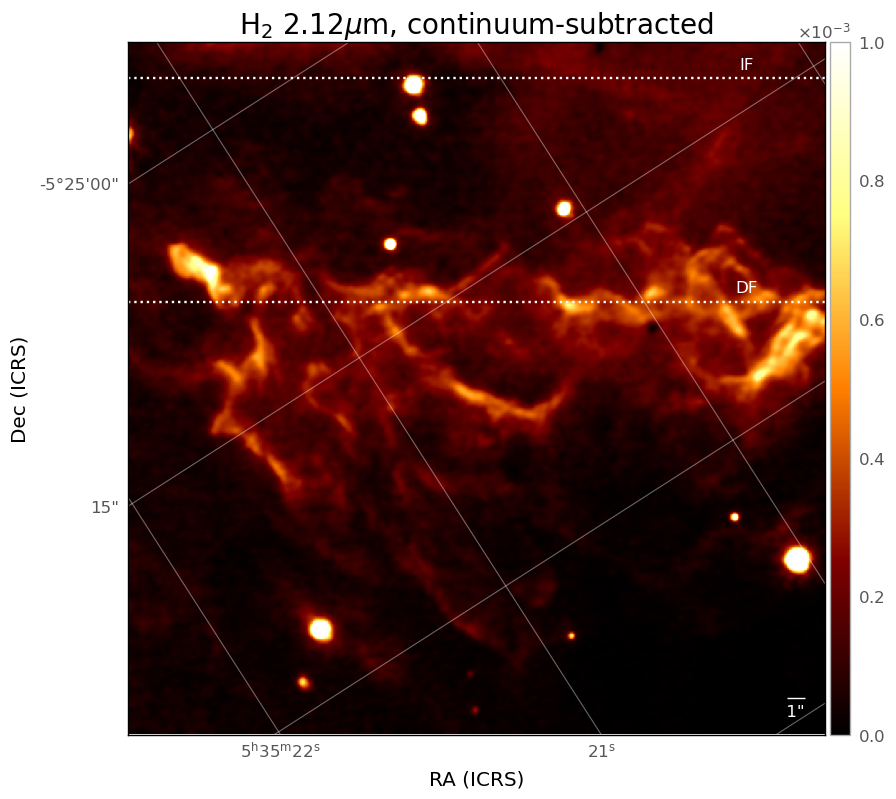}   
    \caption{Same as for Fig.~\ref{fig:Keck-map} for the H$_2$ 1-0 S(1) line emission.
    %Integrated intensity maps of the H$_2$ 1-0 S(1) and continuum-subtracted H$_2$ 1-0 S(1) line emission. Units are erg cm$^{-2}$ s$^{-1}$ sr$^{-1}$ $\mu {\rm m}^{-1}$ for the  H$_2$ map and erg cm$^{-2}$ s$^{-1}$ sr$^{-1}$ for the continuum-subtracted line emission map. The horizontal dashed lines indicate the average position of the ionization and dissociation fronts as determined from the averaged emission cut over the entire FoV presented in the Fig. \ref{fig:Keck-cut}. 
    }
    \label{fig:Keck-map-H2}
\end{figure*}

\begin{figure*}[h!]
    \centering
     \includegraphics[width=0.4\textwidth]{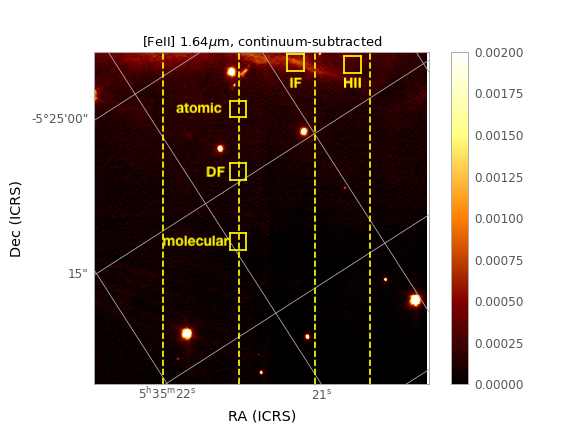}
         \includegraphics[width=0.4\textwidth]{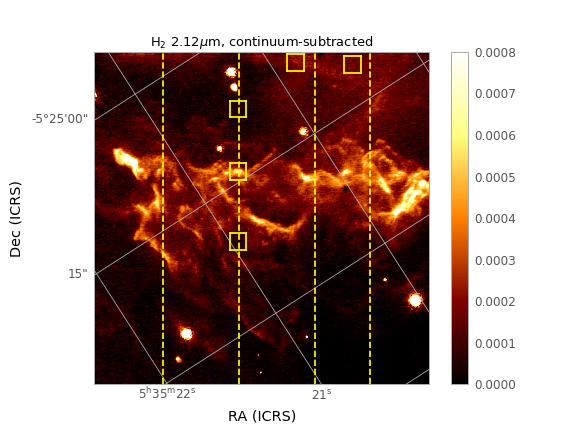}
    \includegraphics[width=0.3\textwidth]{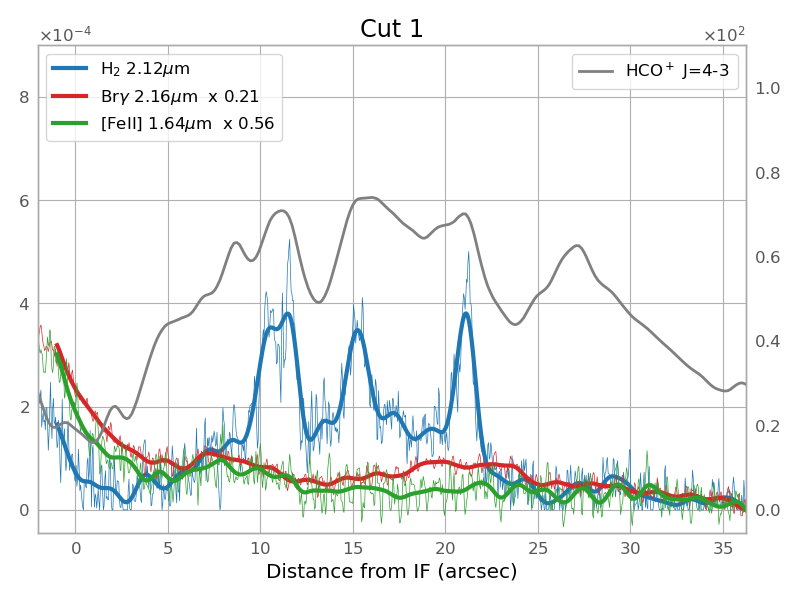}
       \includegraphics[width=0.3\textwidth]{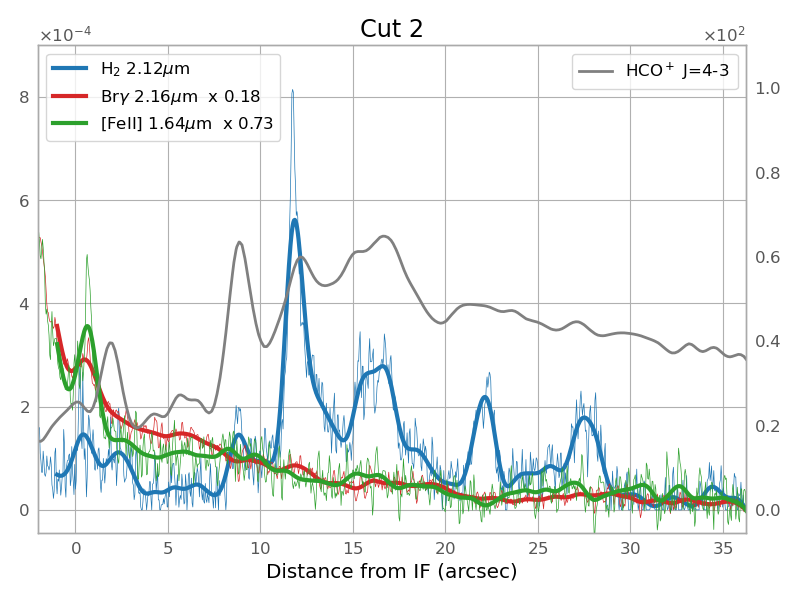}
          \includegraphics[width=0.3\textwidth]{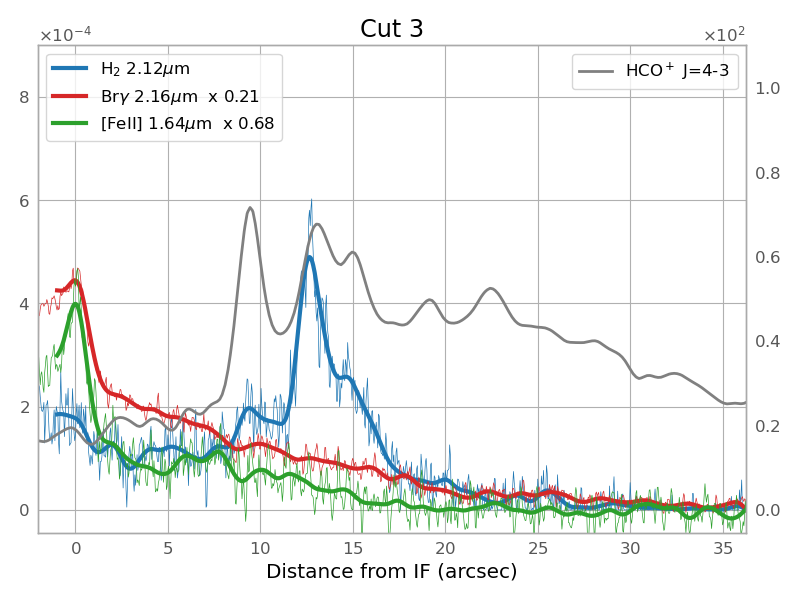}
             \includegraphics[width=0.3\textwidth]{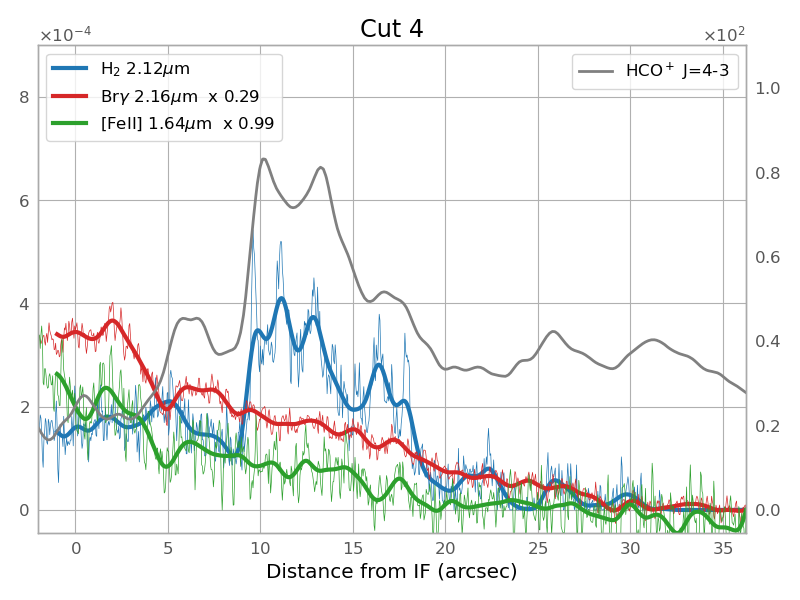}
                         \includegraphics[width=0.3\textwidth]{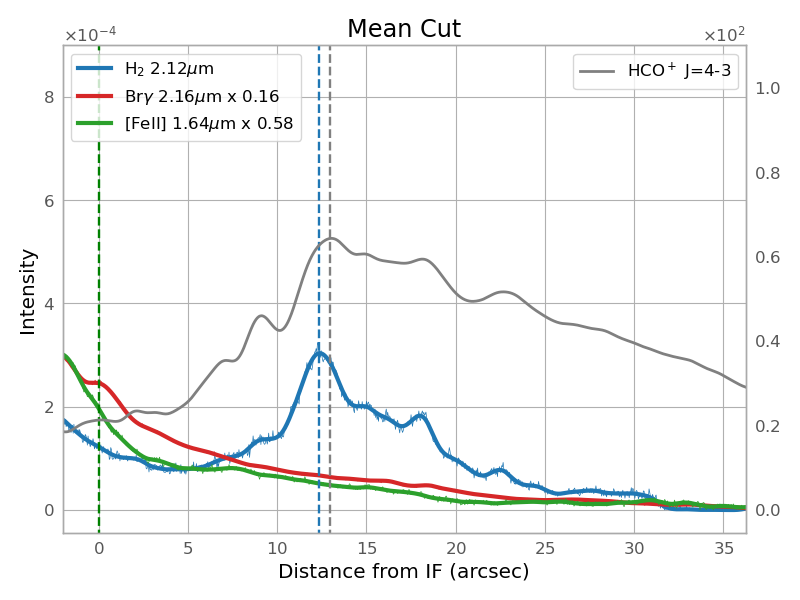}
    \caption{Top row: Superposition of the four cuts perpendicular to the Orion Bar described in Sect.~\ref{sec:spatial-distribution-ionised-gas} on the [FeII] 1.64$\mu$m and H$_2$ 2.12$\mu$m continuum subtracted emission line maps. Cuts 1, 2, 3, 4 are positioned from left to right. The squares area are used to compute the lines intensity in the key regions presented in Table 1. Second and third rows: H$_2$ (blue), Br$\gamma$ (red) and [FeII] (green) line intensity (in erg s$^{-1}$ cm$^{-2}$ sr$^{-1}$) profiles shown as function of the distance from the IF for each cut along with the HCO$^+$ line integrated intensity profile (in K km s$^{-1}$, grey). The intensity profiles of the Br$\gamma$ and [FeII] lines have been scaled by the factors indicated in the legends. The line intensities have not been corrected for dust extinction. Bottom row:  The averaged emission cut calculated by taking the mean over the entire FoV and perpendicular to the Orion Bar.  The vertical dashed lines indicate the average position of the IF, DF and HCO$^+$ peak. 
    %Towards the cuts 2 and 3 with the HCO+ emission peak in between the IF and the H2 emission corresponding to the Background OMC1.}
    }
    \label{fig:Keck-cut}
\end{figure*}

\subsection{Br$\gamma$ and iron lines}
\label{sec:spatial-distribution-ionised-gas}

The maps obtained in the Br$\gamma$ and [FeII] filters
%and Hcont and Kcont filters 
and the continuum-subtracted emission are shown in Fig.~\ref{fig:Keck-map}, while the H$_2$ filter maps are shown in Fig.~\ref{fig:Keck-map-H2}. Most of the structures seen in these lines and the continuum are very similar to those visible in the H$\alpha$ HST map (Fig. \ref{fig:HST}).  The ionization front is clearly seen in the Br$\gamma$ and [FeII] line emission maps.
%The spatial distribution of the Br$\gamma$ and [FeII] lines and the adjacent continuum spatially coincides well for most regions of our maps. 
However, there are some differences between the  Br$\gamma$ and [FeII] line emission morphologies. In particular, in the area surrounding the two identified proplyds near the Bar, 203-506 and 203-504 (%listed in the  HST Atlas of a \cite{Ricci2008}
\citealt{bally2000disks} and located in the yellow squares in Fig.~\ref{fig:HST}): both proplyds emit in [FeII] along with bright streamers, whereas only the proplyd 203-504 shows strong Br$\gamma$ emission as further described in Section~\ref{sec:proplyds}.

The surface brightness profiles measured in the [FeII], Br$\gamma$ and H$_2$ lines along 4 cuts perpendicular to the Bar are shown in Fig.~\ref{fig:Keck-cut}. These cuts use a 3-pixel average corresponding to the typical FWHM of the PSF. Cut 1 goes through the left edge of bar, cut 2 through the position of the JWST/NIRSpec-IFU and MIRI-IFU mosaics of the ERS program \citep[see Fig. 3 in][]{paper_ERS_PASP_2022}, cuts 3 and 4 through the right edge of the bar (see Fig.~\ref{fig:Keck-cut}).
To investigate the stratification of the interfaces from ionized to molecular emission, we also constructed averaged emission cut over the entire FoV and perpendicular to the Orion Bar. As expected for a PDR seen edge-on, we found that the H$_2$ emission is displaced further into the cloud than the Br$\gamma$ and [FeII] line emission. 
A precise determination of the distance between the IF delineated by the [FeII] and Br$\gamma$ lines and the DF delineated by the H$_2$ line can be obtained.  From the averaged emission cuts, we estimated that the offset between the IF and the main peak of the H$_2$ vibrational emission is between 12 and 12.5$''$ (0.024-0.025 pc). %12.34
 Finally, the [FeII] and Br$\gamma$ surface brightness profiles also show extended emission towards the PDR. This emission may originate in the flattened region beyond the bar that is still illuminated directly by the ionizing star \citep{Wen1995, Odell01}.
% some emission in the PDR. This emission may originate from the extended emission from the ionized gas in the line of sight towards the Bar. 

%Comments on the decline of the emission towards the Bar and a flux and non-zero line emission in molecular regions that result from ionized gas contamination along the line of sight. 

\subsection{Vibrationally excited H$_2$ line}
\label{sec:spatial-distribution-molecular-gas}

The maps obtained in the H$_2$ 2.12$\mu$m filter and the continuum-subtracted H$_2$ line emission are shown in Fig.~\ref{fig:Keck-map-H2} and also in Fig.~\ref{fig:Keck-ALMA-map} for comparison with the ALMA HCO$^+$ $J=4-3$ line emission map (see Section \ref{sec:H2-ALMA}). In the H$_2$ line emission map, several bright H$_2$ emission peaks are spatially resolved and show small scale structures.
The H/H$_2$ dissociation front appears highly structured with several ridges and extremely sharp filaments with a width of $1$ to $2''$ (0.002-0.004~pc or 400-800~AU, see Figs. ~\ref{fig:Keck-map-H2} and ~\ref{fig:Keck-cut}). Ridges and filaments run parallel to the dissociation front but a succession of bright sub-structures in H$_2$ is also observed from the edge of the DF towards the molecular region.
The H$_2$ emission peaks appear in an area that starts at about $10''$ from the IF and up to $20$'' and even 25-30$''$ in some places from the IF (see Figs.~\ref{fig:Keck-map-H2} and ~\ref{fig:Keck-cut}).
The bright sub-structures result in several peaks in the H$_2$ brightness profiles shown in Fig.~\ref{fig:Keck-cut} obtained along the cuts perpendicular to the Orion Bar. The peaks are roughly spaced by a few arcseconds ($\sim$ 0.005-0.01~pc or $\sim$1000-2000~AU). 
The surface area occupied by the bright sub-structures (ridges, filaments), i.e. with H$_2$ line intensity $> 2 \times 10^{-4}$ erg s$^{-1}$ cm$^{-2}$ sr$^{-1}$, corresponds to about 40-45\% of the H$_2$ emission zone in the Bar.

Since the H$_2$ emission is very sensitive to both the FUV radiation field and the gas density (as explained below), 
these very narrow and bright sub-structures must be due to dense material directly irradiated. 
Towards more tenuous material, the H$_2$ emission is spatially much more extended and weaker. 
The multiple H$_2$ emission peaks along and across the Bar may be associated with a multitude of small highly irradiated and dense PDRs. 
A single edge-on dense PDR produces one narrow H$_2$ emission peak. 
The slight tilt of the Bar along the line of sight combined with the presence of over-dense structures may explain the observed complex emission distribution, and especially the succession of H$_2$ peaks across the Bar.  

The intensity variations in the different H$_2$ emission peaks ranging from $\sim$2 to $\sim$10 10$^{-4}$ erg s$^{-1}$ cm$^{-2}$ sr$^{-1}$ (Fig.~\ref{fig:Keck-cut}) may result from a combination of effects due to the local gas densities, geometry and dust extinction effects. 
Density effects must important. 
In fact, PDR models %(including both far-ultraviolet-pumping and excitation collisions) 
show that for the conditions prevailing in the Orion Bar, the intensity of the H$_2$ v=1-0 S(1) line is approximately proportional to the gas density
 \citep[e.g.,][]{Burton90}.
  At equilibrium  between the formation of H$_2$ and the photodissociation of H$_2$ by FUV flux, the intensity of the 1-0 S(1) line is proportional to $R_f n_H N(H_0)$ where $R_f$ is the H$_2$ formation rate, $n_H$ is the total hydrogen gas density and $N(H_0)$ the column density of atomic H atoms from the PDR edge. 
For low value of $n_H/G_0$ ($<40$ cm$^{-3}$), which is the case in the atomic zone of the Orion Bar with $G_0\sim10^4$ and $n_H$=10$^4$-10$^5$ cm$^{-3}$, the H$^0$/H$_2$ is driven by dust opacity and N(H$^0$) is a constant equal to a few 10$^{21}$ cm$^{-2}$. The intensity is thus linearly proportional to the density. % and nearly independent of G$_0$. 
Moreover, the H$_2$ and HCO$^+$ J=4-3 emission show a remarkable similar spatial distribution as shown in Fig.~\ref{fig:Keck-ALMA-map} and further described in  Sect. \ref{sec:H2-ALMA}. 
This rotational line is a good indicator of dense gas \citep{Goico16}. 
 
The dense sub-structures seen in H$_2$ may be surrounded by a lower-density gas component producing an extended emission. 
In the H$_2$ data, a more widespread and extended emission (with an intensity of about 2~10$^{-4}$ erg s$^{-1}$ cm$^{-2}$ sr$^{-1}$) seems to be in fact observed at the photodissociation front at about 10-15$''$ (0.02-0.03~pc) and up to 20-30$''$ (0.04-0.06~pc) from the IF. 
Moreover, the H$_2$ map and brightness profiles show some emission in front of the DF (see Fig.~\ref{fig:Keck-cut} and Table~\ref{tab_intensity}). 
Part of this emission likely originates from the surface of the Orion molecular cloud-1 (OMC1). 
This surface is perpendicular to the line of sight and is illuminated by the Trapezium cluster, making it a face-on PDR. 
The emission from this background face-on PDR was also observed with Herschel in other PDR tracers, especially in high-J CO lines \citep{Parikka2018} and [OI] 63 and 145 $\mu$m and [CII] 158 $\mu$m \citep{Bernard-Salas12}. An increase in the H$_2$ emission at the ionisation front is also observed. This was also visible in the previous data of \cite{Walmsley00} and  could % without obvious explanation. %This emission can 
 originate from the background molecular cloud.
%Javier : Any explanation/implication? More H2 than expected or the little H2 we observe is very efficiently pumped?

%In the area of the dissociation front, the H$_2$ emission comes (i) on the one hand from the irradiated dense substructures (making the narrow emission filaments) and (ii) on the other hand from a spatially extended emission due to the less dense gas in the bar and in which is immersed the sub-structures. %, as well as, the background face-on PDR.  This is in agreement with several previous studies as \citep{Goico16,Joblin18} that claimed that the Orion Bar is composed by very small ($\sim$1-2'') dense highly irradiated PDRs with differing local density immersed in a less dense environment spatially extended ($\sim$10-20ÕÕ).

\subsection{Comparison with previous H$_2$ observations}

Our Keck observations convolved at the angular resolution of $\sim$1'' of the previous SOFI/NTT observations by \cite{Walmsley00} show good agreement in terms of line distribution and intensity (see Table~\ref{tab_intensity}). It can be noted that in the SOFI/NTT observations, the brightest sub-structures were already observed but not spatially resolved. The TEXES observations of the pure rotational H$_2$ 0-0 S(1) and S(2) lines by \cite{Allers05}, with a resolution of $\sim$2'' cover the south-west part of our map. The comparison of these observations with the H$_2$ 1-0 S(1) line map  obtained previously by \cite{vanderWerf96} show that the spatial distribution of the pure rotational lines and the vibrationally excited line agree in remarkable detail (see Fig.~2 in \citealt{Allers05}). This is predicted by high pressure models \citep{Allers05,Joblin18},
where a separation between
the H$_2$ lines is expected to be very small ($\lesssim$0.5$''$). Thus, one would expect to observe an overall spatial coincidence on the whole Keck map. Nevertheless, detailed variations are expected because the pure rotational lines that result from collisional excitation have a different dependence on the local physical conditions. Future JWST-ERS observations \citep{paper_ERS_PASP_2022} will allow such an investigation in the parts of our maps observed with the IFU-MIRI spectroscopy.

%H2 emission detected at H/H2 transition : diffuse + structured emission ? filaments = micro PDRs everywhere. Excess of H2 emission at the transition regions between diffuse and dense medium. In the PAH emission = emitting both in the atomic+warm molecular media = diffuse + structured medium = smooth emission => not highly structured. This may reflect differing local gas densities. 
%H$_2$ line emission maxima are resolved at roughly periodic separation. Towards the cut 2, H$_2$ line emission maxima are resolved at roughly periodic separation of about 5"

\subsection{Comparison between H$_2$ and HCO$^+$ from ALMA}
\label{sec:H2-ALMA}
In this section, the distribution and correlation of the H$_2$ line emission with the ALMA observation of the HCO$^+$ J=4-3 line are analysed. 
 Considering the position of the compact sources (e.g. proplyds 203-506), the ALMA observations showed an offset to our Keck observations in coordinates of R.A.=-0.15" and DEC.=0.75". By correcting this offset, a good overlap of the position of the compact sources and of the Bar is obtained.

\begin{figure*}[h!]
    \centering
        \includegraphics[width=0.8\textwidth]{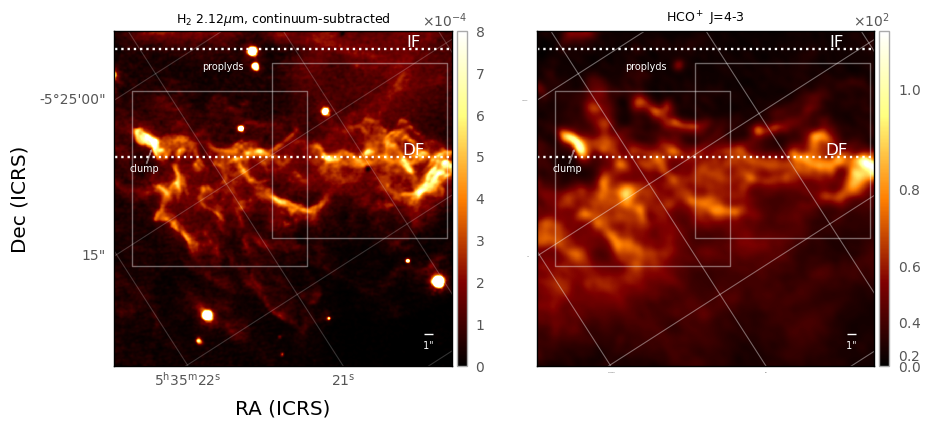}
                \includegraphics[width=0.8\textwidth]{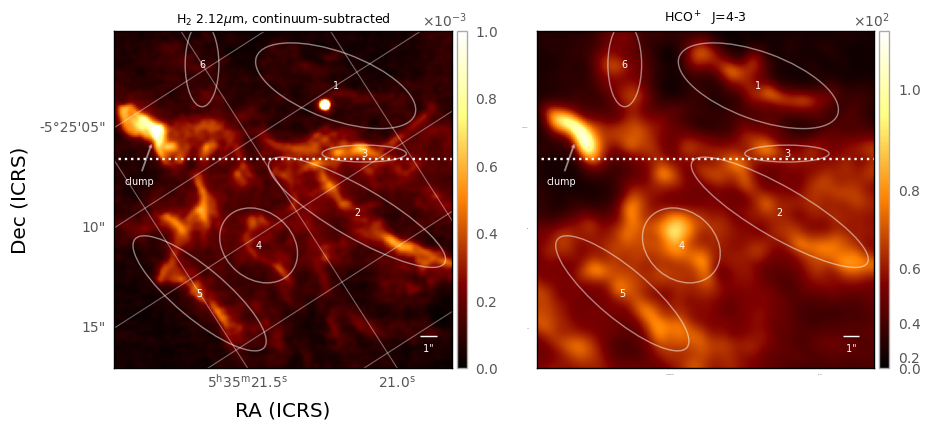}
                        \includegraphics[width=0.8\textwidth]{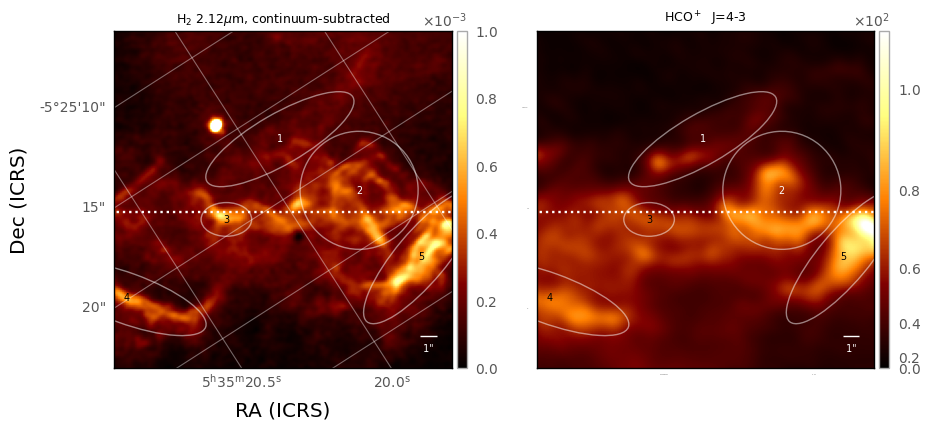}
    \caption{Left panels: Integrated intensity maps (continuum-subtracted) of the H$_2$ 1-0 S(1) line emission in erg cm$^{-2}$ s$^{-1}$ sr$^{-1}$. Right panels: Integrated intensity maps of the HCO$^+$ J=4-3 lines in K km s$^{-1}$.}
    \label{fig:Keck-ALMA-map}
\end{figure*}

In Fig.~\ref{fig:Keck-ALMA-map}, we compare the maps obtained in the H$_2$ 1-0 S(1) and HCO$^+$ J=4-3 lines across the same field of view. The middle and bottom figures zoom into two parts of the map to be able to compare in detail the spatial distribution of the emission from each of the detected sub-structures. Most of the sub-structures are common to both maps and show a very similar distribution. % (see the sub-structures denoted 2, 3, 4, 5 in the middle panels and the sub-structures denoted 2, 4, 5 in the bottom panels of the figure).
Due to a high-dipole moment and a high critical density, the HCO$^+$ 4-3 line (n$_{critical}$ of a few 10$^6$ cm$^{-3}$) is a good proxy of the gas density \citep{Goico16}. Its integrated intensity is roughly proportional to the density in the $n_H=10^4-10^6$ cm$^{-3}$ range. %Excitation models show that 
The average density that reproduces the mean HCO$^+$  towards the dissociation front (at about 15$''$ from the IF) is about $n_H=0.5-1.5~10^6$ cm$^{-3}$.
%ALMA HCO+ (J = 4 ? 3) observations close to the DF indicate  densities of n = 106 and sizes of 0.004 pc. 
Thus, some of the densest portions of the Bar lie along the dissociation front. We note that the small H$_2$ and HCO$^+$ structures discussed here localized at the DF are shifted by a about $\sim$20'' relative to the bigger (5"-10") condensations seen more inside the molecular cloud \citep{YoungOwl00,Lis03}.
%HCO+ line emission maxima are resolved at roughly periodic separation. Wave front space by 5" ?

Although there is a very good spatial coincidence between the H$_2$ and HCO$^+$ sub-structures, the H$_2$ line emission decreases faster in the Bar than the HCO$^+$ line emission (see Fig.~\ref{fig:Keck-cut}). This can be explained as follows. 
Firstly, the vibrationally excited H$_2$ line is more sensitive to the FUV field flux and the most intense H$_2$ emission must come from the transition region between the diffuse and very dense medium where the FUV radiation is not yet attenuated. The FUV shielding produced by the ridge of high-density sub-structures may significantly decrease the vibrational excited H$_2$ in the deeper regions. Secondly, extinction due to dust in the Orion Bar and to foreground dust may affect the apparent morphology of the near-IR H$_2$ images. Small-scale dust extinction differences could in fact  result in morphological differences between the near-IR and millimeter-wave images. Dust extinction effects on the line intensities is discussed in Sect. \ref{sec:intensity-extinction}. 

The overall remarkable spatial coincidence between the H$_2$ and HCO$^+$ line emission shows that they both come from high-densities but also indicates their strong chemical link. Detection of both bright HCO$^+$ and CO emission by ALMA towards the H$_2$ vibrational emission layers (Fig.~2c in \citealt{Goico16}) suggests that the C$^+$/CO transition nearly coincides with the H/H$_2$ transition. 
This fact was predicted in \cite{Joblin18}, where the best fitting Orion Bar model has a separation between these transitions of $\sim$0.25$''$. 
%(although the scale of the atomic region was to compact by a factor of ~3).
%The HCO+(J = 4 ? 3), high-J CO (Parikka et al. 2018), and H?2(v = 1 ? 0) are nearly coincident, suggesting a merging of the H2 and CO dissociation layers (Goicoechea et al. 2016, Kirsanova & Wiebe 2019). 
We obtained a precise determination of the average offset between the peak of the H$_2$ vibrational emission (delineating the H/H$_2$ transition) and the edge of the observed HCO$^+$ emission (delineating the C$^+$/C/CO transition).  From the averaged emission cuts, we estimated that this average offset is less than $1"$, about $\sim$0.6" (or 0.0012~pc). However considering that the angular resolution of the ALMA data is $\sim$1$''$, higher angular resolution observations are required to measure the exact value of this offset.
%HCO+ line emission maxima are resolved at roughly periodic separation. Wave front space by 5" ?
%However, as underlined in \citealt{wolfire22} this may also be due to high densities and resulting small scale sizes, with vibrationally excited H$_2$ driving the carbon chemistry to produce both CO and HCO$^+$. 
The presence of a large quantity of irradiated H$_2$ may have important consequences for the chemical structure. Due to the high densities and enhanced H$_2$ gas heating via formation and collisional de-excitation of FUV pumped levels, the chemistry is triggered by high temperature and FUV pumped levels. Endothermic reactions and reactions with energy barriers become faster \citep{Agundez10,Nagy13}. Reactions of H$_2$ with abundant atoms and ions, e.g. C$^+$, shift the carbon bearing molecular gas towards the cloud edge.

We note that although a spatial coincidence is observed between the H$_2$ and HCO$^+$ over-dense sub-structures, a spatial shift of the emission peak is measured for several sub-structures (e.g., sub-structure 3 in the middle panels and sub-structures 2, 3 in the bottom panels of Fig.~\ref{fig:Keck-ALMA-map}). This may essentially result from the fact that the two emission lines vary differently with the local conditions (FUV field flux and gas density) and that the H$_2$ near-IR line  is affected by dust extinction. The most important differences in the spatial distribution of the H$_2$ and HCO$^+$  emission are found in the atomic region. In fact, the brighter structures seen in the atomic regions in HCO$^+$ (like globulettes) are faintly visible in H$_2$ ro-vibrational emission (e.g., the sub-structures denoted 1 and 6 in the middle panels and the sub-structure denoted 1 in the bottom panels of Fig.~\ref{fig:Keck-ALMA-map}). These structures are probably located in the background or foreground. %Velocity contraints from ALMA data ?
Indeed, the emission velocity of these HCO$^+$ structures ($v_{LSR}$= 8-9 km s$^{-1}$) is more consistent with emission from the background OMC1 than from the Bar ($v_{LSR}$= 10.5 km s$^{-1}$). 
Some extended H$_2$ line emission seen in the atomic zone to the south-west part of the Bar (see bottom panel of Fig.~\ref{fig:Keck-ALMA-map}) has also no  counterpart in the HCO$^+$ or CO line emission maps. Finally, plume-like CO features seen with ALMA in the atomic zone by \cite{Goico16} which may be advected from the surface of the molecular cloud have no evident correspondences with the H$_2$ emission structures observed by Keck/NIRC2.

 %The few differences in the spatial distribution may result from the fact that HCO+ is sensitive to higher gas density? Javier could provide a chemical tool?
 
\subsection{Proplyds and collapsing protostars}
\label{sec:proplyds}

We are able to spatially resolve the FUV irradiated Orion Bar PDR edge but also additional compact sources corresponding to over-dense irradiated sub-structures such as proplyds near the IF, and perhaps collapsing protostars near the DF. In the following, we briefly describe the observations towards these sources.

\subsubsection{Proplyds}

\begin{figure*}[h]
    \centering
         \includegraphics[width=1.0\textwidth]{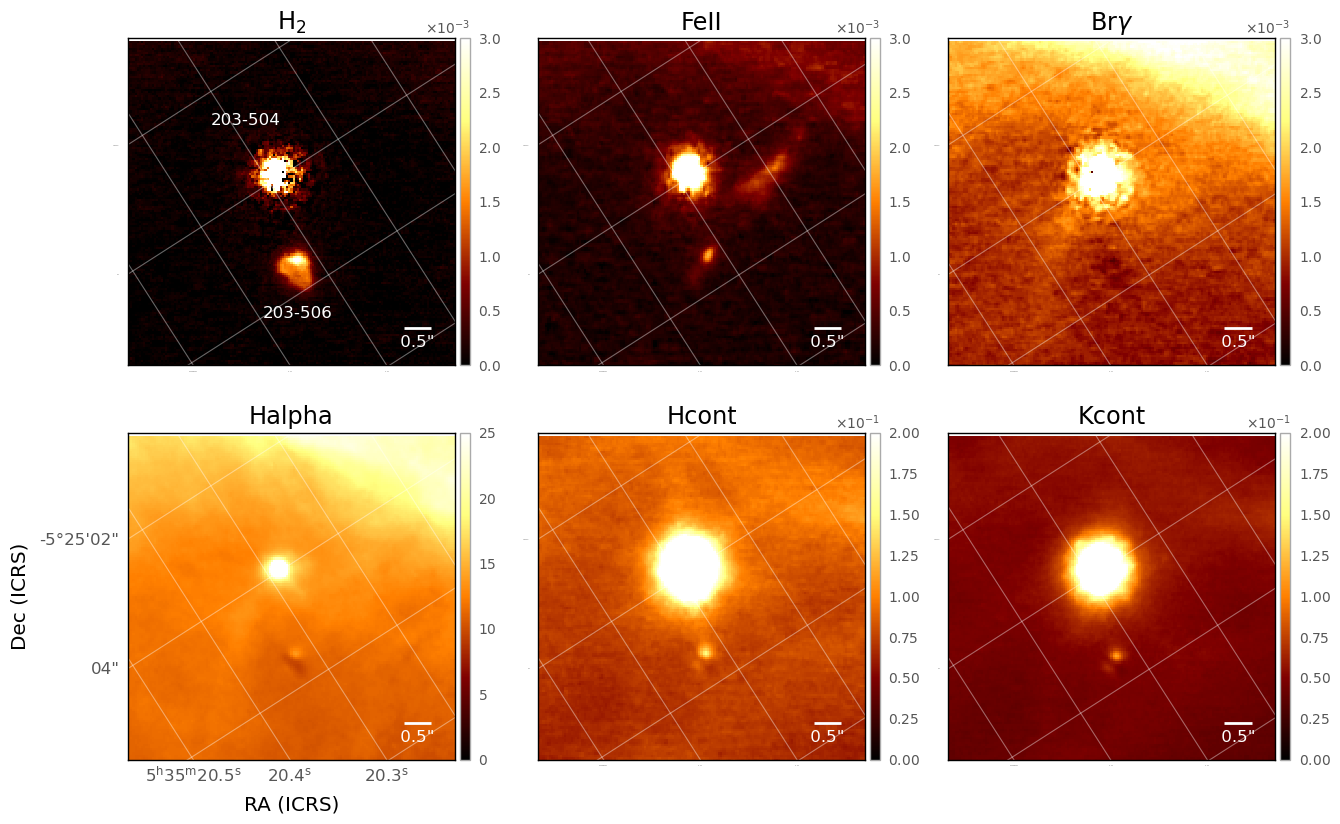}
    \caption{Zoom on proplyds in the H$_2$ 2.12$\mu$m, [FeII] 1.64$\mu$m and Br$\gamma$ 2.16$\mu$m lines (continuum substracted), as well as, in the Hcont and Kcont filters with in addition the H$\alpha$ emission using the HST observations from \texttt{https://archive.stsci.edu/prepds/orion/acs\_displaypage.html} which was adjusted in positions.}
    \label{fig:proplyds}
\end{figure*}

Fig. \ref{fig:proplyds} shows proplyds 203-504 and 203-506 \citep{bally2000disks} which are present in the FoV of our Keck/NIRC2 observations.
The star of 203-504 is detected in all filters, however the disk/envelope is not clearly identifiable in these images.
%Origin: saturation and/or PSF fluctuation in H2 and cont..
The 203-506 proplyd is clearly visible in emission in both the H$_2$ and [FeII] lines, whereas it is not detected in the Br$\gamma$ line.

The H$_2$ emission appears to have a similar morphology to the [OI] 6300 \AA~line observed by \citet{bally2000disks}. This is expected since both lines trace warm neutral gas ($\sim 1000$ K) near the dissociation front, inside the dense PDR that is created at the disk surface \citep{Champion17}.
The [FeII] line shows a very different morphology, with an elongated structure perpendicular to the proplyd disk, and is reminiscent of a jet, such as those identified for proplyds by \citet{Bally2015}. The [FeII] emission would then be associated with irradiated shocked gas.  
The absence of Br$\gamma$ emission could have different explanations.
The coming PDR4alls JWST data should bring additional clues to better understand it.

%could be explained by the fact that 203-506 is supposed to be embedded in neutral atomic gas \citep{Champion17},  as it is located behind the IF of the Orion Bar (with respect to the Trapezium cluster). However, some aspects remain puzzling. One could expect that in order to appear as dark in emission line, the proplyd should have an emission background

%It is also interesting to note the similarity in between the H$_2$ and [OI]  spatial distributions, that are likely tracing the shocked region in between the proplyd envelop and jet. The latter seems to be well characterized by the FeII emission, which displays an emission direction perpendicular to the Orion Bar and the brightest shocked surface traced H$_2$. 

\subsubsection{Over-dense sub-structures}

Numerous over-dense small structures (1-2'') appear in the H$_2$ and HCO$^+$ emission that could be self-gravitating, or transient, turbulently compressed features or compressed by FUV photoevaporation \citep{Gorti02,Tremblin2012}.
%could have been compressed by FUV photoevaporation 
%The over-dense small structures (1-2'') seen in H$_2$ and HCO$^+$ emission are displaced relative to the bigger (5"-10") condensations previously detected inside the molecular cloud \citep{YoungOwl00}. 
Recently, \cite{Rollig2022arXiv} showed that the intensity levels of the HCO$^+$ emission?map as well as its spatial distribution is consistent with predictions from nonstationary, clumpy PDR model ensembles.
Whether they can become star-forming clumps (for example, by merging into massive clumps) is uncertain. 
% Hosokawa, T. & Inutsuka, S.-i. Dynamical expansion of ionization and dissociation front around a massive star. II: on the generality of triggered star formation. Astrophys. J. 646, 240-257 (2006).
Gravitational collapse appears not apparent from their density distribution \citep[no high-density power-law tail, ][]{Goico16}. %Indeed, their estimated masses are much lower than the mass needed to make them gravitationally unstable. 
In fact, the mass of a cylinder with a density $n_H \sim 10^6$ cm$^{-3}$, a width of 1" and a length of 2"-6" is $<0.005$~M$_{\odot}$ \citep{Goico16}, which is much lower than the virial and critical masses needed to make them gravitationally unstable (approximately 5$M_{\odot}$ from the inferred gas temperature, density, and velocity dispersion,  \citealt{Inutsuka1997}). 
However, the very bright substructure located in the northeast of our map and denoted clump in Fig.~\ref{fig:Keck-ALMA-map} shows a very particular structure (with a size of $\sim$1000~AU), likely a  YSO candidate (Goicoechea et al. in
 prep).
%which could be a potential streamer  and be the evidence of collapsing.
% This potential streamer of infalling flows can be traced by warm irradiated gas feeding a protostar.
%ALMA observations revealed evidence for small  accretion streams (less than 1000 au) originating from within the dense core of this plausible YSO \citep[e.g.,]{Yen2014} while NOEMA observations of a Class 0 object reveal a previously unseen large-scale ($\sim$10,000 AU) streamer of fresh gas from the surrounding dense core down to the disk scales \citep{Pineda2020}.

\section{Line intensities and extinction correction}
\label{sec:intensity-extinction}

\subsection{Line intensities in the key zones}

In Table~\ref{tab_intensity},  we give the intensities of the [FeII], Br$\gamma$, and H$_2$ lines measured towards the ionised region, IF, PDR atomic region, DF, and from the molecular region. 
The line intensities of each of these characteristic regions have been estimated in the 2" by 2" square areas shown in Fig.~\ref{fig:Keck-cut} (top panels). 
The areas for the atomic, DF, and molecular zones which are located at about $5.5''$, $12.5''$, $21''$ from the IF, are positioned within the FoV of the NIRSpec-IFU and MIRI-IFU mosaics (see Fig.~3  of \citealt{paper_ERS_PASP_2022}) which are along the direction of cut 2 in Fig.~\ref{fig:Keck-cut}. On the other hand, the areas chosen for the ionized and IF regions are not aligned along the JWST mosaics (i.e. cut 2) because the Keck maps do not entirely cover these ionized gas and IF regions along the JWST mosaic (i.e. cut 2). We therefore chose areas for the ionized gas and IF regions further south-west (towards the direction of cuts 3 and 4, see Fig.~\ref{fig:Keck-cut}).

Comparison with previous measurements of the intensities of these lines with SOFI/NTT by \cite{Walmsley00} in the ionized, IF, and DF regions %corresponding to their A, B, C regions (see their Fig. 1 and Table 1) 
are given in Table \ref{tab_intensity}.
The measurements are in good agreement within a factor of about 2 for the [FeII] and Br$\gamma$ lines at the IF and the H$_2$ line at the DF. However, we find some larger differences for Br$\gamma$ in the HII and DF regions and H$_2$ in the HII and IF regions. This might be mostly  due to the fact that the position regions of \cite{Walmsley00} (shown as A, B, C in their Fig.~1) do not correspond to our positions.  Their HII and IF regions are outside of the Keck/NIRC2 field-of-view. % correspond to our positions. 
%The median H$_2$ line emission on the average cut is about 3 10$^{-4}$ erg cm$^{-2}$ s$^{-1}$ sr$^{-1}$ in agreement with \cite{Walmsley00}. But on the brightest filaments the H$_2$ emission can reach values of about 5 to 10 10$^{-4}$ erg cm$^{-2}$ s$^{-1}$ sr$^{-1}$.  
%The observed [FeII] line intensity  is $XX\times10^{-4}$\,erg/s/cm$^2$/sr at the ionization front and $XX\times10^{-4}$\,erg/s/cm$^2$/sr in the atomic PDR. The observed [FeII] line intensity with SOFI/NTT is $6\times10^{-4}$\,erg/s/cm$^2$/sr at the ionization front and $1.2\times10^{-4}$\,erg/s/cm$^2$/sr in the atomic PDR.

\subsection{Extinction correction}
%We estimated dust extinction corrections on the lines intensity. 
Corrections for extinction due to the foreground dust and internal dust in the Bar itself of the line intensities were estimated and reported in Table \ref{tab_intensity}.
The foreground visual extinction $A_V$ produced by the Veil was estimated to be not greater than about 2 mag by \cite{Odell2000} using a radio-to-optical surface brightness comparison.
Using flux measurements of the Balmer lines (H$\alpha$ to H$\beta$ flux ratio), \cite{weilbacher2015} estimated that the foreground extinction is about $A_V \sim$ 1.3 mag (or $A_K \sim$ 0.15 mag) towards the ionized gas in the line of sight of the Orion Bar (see their Fig.~15). The foreground dust structure appears smooth and devoid of significant substructure. 
By this dust screen, the line emission will be attenuated by $e^{-\tau _{\lambda}}$ with $\tau_{\lambda}=0.921 ~A_{\lambda}$ the optical depth at the wavelength of the line. To adopt a dust reddening  appropriate to Orion, we follow \cite{goicoechea15} and used the extinction curve of \cite{Cardelli89} as refined by \cite{blagrave2007} with $R_V =A_V /E(B-V)=$5.5. % and $A_{\lambda}/A_v$ values in Cardelli et al. (1989).  $E(B-V)= A_B - A_V$ is the reddening factor, and $R_V$ is a dimensionless parameter that characterizes the slope of the extinction curve.
The reddening factors are 0.216, 0.143 and 0.138 at 1.64, 2.12 and 2.16 $\mu$m respectively.
%for the three gas lines considered here and the K band are $A_{1.64 \mu m}/A_V=0.216$, $A_{2.16 \mu m}/A_V=0.138$, $A_{2.12 \mu m}/A_V=0.143$ and $A_{K=2.2 \mu m}/A_V=0.135$.
Given this degree of foreground extinction, we estimated that the [FeII], Br $\gamma$, and H$_2$ lines are approximately 23\%, 15\%, and 16\% brighter respectively if foreground extinction correction is taken into account.
%The line intensities corrected from this foreground extinction are presented in Table \ref{tab_intensity}. 

For the H$_2$ line in the PDR, 
internal extinction due to the dust in the Bar itself may significantly attenuate the emission. An additional magnitude of extinction due to dust in the Bar between the ionized gas and the region of 
excited H$_2$ may result in a stronger increase in line intensity. In order to estimate this extinction, we assumed that the column density of gas along the line of sight is equal to the column density between the IF and the region of 
excited H$_2$.
Thus, we used the density profile towards the Bar as constrained from comparison between Spitzer and Herschel observations and a radiative transfer code with the THEMIS dust model \citep{Schirmer2022}. 
The corresponding calculated column density is 1.5~10$^{21}$ H cm$^{-2}$ from the IF to the beginning of the H$_2$ emission (at 10'' or 0.02~pc from the IF) and 10~10$^{21}$ H cm$^{-2}$ from the IF to the end of the H$_2$ emission (at 20'' or 0.04~pc from the IF). %For simplicity, it is considered that the density profile of the dust in the atomic region seen from the edge or from the front is similar. 
Adopting the same dust reddening as described before from \cite{Cardelli89} and $A_V /N_H=3.5 \times 10^{-22}$ mag cm$^{-2}$, we estimated an extinction of $A_V=1$ to 4 or $A_K=0.13$ to 0.54. If this internal dust extinction correction is taken into account and a correction factor of  $e^{-\tau_{\lambda}}$ is applied, the H$_2$ line is approximately 12 to 40\% brighter. %Variation of about 28\% on the H$_2$ line intensity can thus be expected due to the value range of the column density. 
Regarding the assumed extinction law, the Cardelli extinction curve with $R_V=$5.5 is similar to that calculated by the THEMIS dust model in the Orion bar with nano-grains depletion \citep{Schirmer2022}. 
For this model, the reddening factors are 0.224, 0.139 and 0.134 at 1.64, 2.12 and 2.16 $\mu$m respectively.
%For this model, the reddening factor for the three gas lines considered here and the K band are $A_{1.64 \mu m}/A_V=0.224$, $A_{2.16 \mu m}/A_V=0.134$, $A_{2.12 \mu m}/A_V=0.139$ and $A_{K=2.2 \mu m}/A_V=0.13$.
The correction factor on the line intensities assuming their extinction law will differ by less than 3\%. 

We must underline that if we assume that the gas and the dust are well mixed in the H$_2$ emission zone, one must apply a correction factor of 
(1-$e^{-\tau_{\lambda}})/\tau_{\lambda}$ \citep[e.g.][]{thronson1990}. Assuming a visual extinction of $A_V=3$ for the total H$_2$ emission zone (including the succession of H$_2$ peaks) as calculated using the dust density model, we computed the corresponding correction factor. This visual extinction is equivalent to consider a gas density of 1-1.5 10$^5$ cm$^{-3}$ and a length of 0.02~pc (as the width of the H$_2$ emission zone in Fig.~\ref{fig:Keck-cut}). This visual extinction corresponding to a large column density of UV pumped H$_2$ can be considered as an upper limit . With these assumptions, the line is approximately 18\% brighter after extinction correction in the H$_2$ emission zone. Considering the increase of 12\% due to the extinction correction of the atomic zone from the IF to the beginning of the H$_2$ emission zone, the line will thus be 30\% brighter. This is relatively close to what we found previously assuming no mixing which is of 40\% (due to the internal extinction correction from the IF to the end of the H$_2$ emission zone). 
% (it corresponds to an internal extinction assuming no mixing of $A_K=0.37$).

In Table~\ref{tab_intensity} and for the comparison to the model predictions in the following section, we assumed that the H$_2$ line will be in total 56\% brighter after correcting for foreground (16\%) and internal extinction (40\%) in the Bar. 
%This is in agreement with the total extinction correction derived by \cite{kaplan2021}. %Moreover, the assumption of well-mixed gas and dust is not obvious considering the fact that the observed H$_2$ emission comes mostly from the dense sub-structures and not the extended less-dense medium.
Our total (foreground and internal) extinction corrections for H$_2$ are thus about $A_K =0.17 + 0.54 = 0.71$ which is very close to the one derived by \cite{Kaplan2021} of $A_K = 0.72 \pm 0.1$.
These authors used an effective way to measure extinction which consists of 
comparing the observed to theoretical H$_2$ ro-vibrational line flux ratios from pairs of lines arising from the same upper level that are widely separated in wavelength. Their errors on $A_K$ were estimated from the assumptions made on the color correction. % that considered $A_{\lambda} = \lambda ^{\beta}$ with $\beta$=1.6 to 2.   
Their IGRINS observation at high spectral resolution were made along a slit positioned across the Bar and centered on at the southwest of the Bar (corresponding to the position of the sub-structure 5 in the bottom zoom of Fig.~\ref{fig:Keck-ALMA-map}). %$\alpha$=05h35m19.73s, $\delta$=-05d25m26.7s (J2000).  This slit center position is at the southwest of the Bar and is localized at the position of the sub-structure 5 in the bottom zoom figure \ref{fig:Keck-ALMA-map}). 
However, we must underline that their extinction measurement is lower than the values derived by \cite{luhman1998} also using H$_2$ ro-vibrational transitions with common upper levels but observed a long a slit  passing roughly through the center of our Keck maps. %approximately from $\alpha$=5h35m19.5s and $\delta$=-5d24m53s (2000) 
\cite{luhman1998} found averaged values of  $A_K$=2.6$\pm$0.7 and 2.3$\pm$0.8  respectively for the H$_2$ peak and beyond.
%the H$_2$ peak (Bar2) and the weaker H$_2$ emission beyond the bar (Bar1).
The H$_2$ line will be approximately 90\% brighter if an extinction correction of $A_K=2.6$ (or $A_V=19$) is taken into account.
The internal extinction might be variable depending on the sightline and the density of the region crossed.
From the radio and NIR H$_2$ line maps, \cite{Walmsley00} suggest in fact that extinction can vary rapidly as a function of position in the Bar. Specifically, precise spatial estimates of the internal PDR extinction are needed. This will be possible with the high angular resolution near-IR line maps that will be obtained with JWST's NIRCAM and NIRSpec IFUs. This will constrain in detail how dust extinction affects the apparent morphology of the H$_2$ line emission. 

In summary, we estimated that the [FeII], Br $\gamma$, and H$_2$ lines will be $\sim$23\%, 15\%, and 16\% brighter if foreground extinction correction is taken into account. For the H$_2$ line in the PDR, 
internal extinction due to the Bar itself may also significantly attenuate the emission and the line could be about $\sim$56\% brighter in total. However, the extinction might vary in the Bar due to due to density variations along the line of sight. % and results in increasing or decreasing that correction. 

%In Table \ref{tab_intensity}, the H$_2$ line intensity at the DF and beyond the bar have been corrected from foreground and internal extinction assuming $A_K=0.7$.
%g(?p,?f) = exp(??f) * (1 ? exp??)/?
%The corrected lines intensity are compared to the model predictions in the following section.

\begin{table*}
\caption{Observed and predicted line intensities in erg cm$^{-2}$ s$^{-1}$ sr$^{-1}$ in the ionized, IF, atomic, DF, and molecular regions. 
For Keck observations, the median and standard deviation of the intensities observed in the 2$''$ by 2$''$ square areas shown in Fig.~\ref{fig:Keck-cut} are given. The intensity peak of the lines is also given. The observed Keck intensities have been corrected for dust extinction and the correction factor applied is given in \% (in the row "Corr. Extinction"). Measurements of the intensities of these lines with SOFI/NTT by \cite{Walmsley00} in the ionized, IF, and DF regions (corresponding to their A, B, and C regions in their Table 1) are also reported.}   
\label{tab_intensity}      
\centering        
\begin{tabular}{ c c c c c c c c c}     % 8 columns
\hline 
\hline
Line & $\lambda$ ($\mu$m) & $A$ (s$^{-1}$) & $E_u$ (K)  & & & Intensities  &  &  \\
 \hline
& &  & & ionized & IF & atomic & DF & molecular  \\
\hline
[FeII]  & 1.64440 & 0.006 & 11445.9 &  &   &   &  &   \\
 &  &  &  &  &   &   &  &   \\
% &  &  & Obs. Keck  & 2.9$\pm$0.6 10$^{-4}$  &  5.7$\pm$1.3 10$^{-4}$ & 1.4$\pm$0.5 10$^{-4}$ &  0.9$\pm$0.5 10$^{-4}$ &  0.6$\pm$0.5 10$^{-4}$\\
%&   &  & Max. Intensity &  4.9 10$^{-4}$  &   1 10$^{-3}$ & &  & \\
 &  &  & Obs. Keck  & 3.6$\pm$0.7 10$^{-4}$ &  7.0$\pm$1.6 10$^{-4}$  & 1.7$\pm$0.6 10$^{-4}$  &  1.1$\pm$0.6 10$^{-4}$  &  0.7$\pm$0.6 10$^{-4}$ \\
&   &  & Max. Intensity &  6 10$^{-4}$  &   1.2 10$^{-3}$ & &  & \\
   &  &   & Corr. Extinction  & +23\%& +23\%& +23\% & +23\%& +23\%\\
   & &  & Obs. NTT & 4.7 10$^{-4}$    & 6.0 10$^{-4}$ & & 1.2 10$^{-4}$  &  \\
   & &  & Cloudy model & 1.9 10$^{-7}$    &        6 10$^{-4}$
   & - & - & - \\
  % 2.21e-07            7.51e-04
%FeII
%HII : Median : 2.87e-04  -- std error: 6.43e-05 --25% error : 7.18e-05 -- Max  : 4.92e-04
%IF : Median : 5.67e-04  -- std error: 1.32e-04 --25% error : 1.42e-04 -- Max  : 1.03e-03
%atomic : Median : 1.40e-04  -- std error: 5.22e-05 --25% error : 3.51e-05 -- Max  : 3.63e-04
%DF : Median : 9.37e-05  -- std error: 5.04e-05 --25% error : 2.34e-05 -- Max  : 2.78e-04
%molecular : Median : 5.91e-05  -- std error: 4.64e-05 --25% error : 1.48e-05 -- Max  : 2.29e-04
\hline
H$_2$  &  2.12183  &  3.47 10$^{-7}$ &  6951.3   &  &    &  & &  \\
%  &   &   & Obs. Keck & 1.6$\pm$0.5 10$^{-4}$ &  1.9$\pm$0.6 10$^{-4}$   & 0.4$\pm$0.4 10$^{-4}$  & 3$\pm$1.5 10$^{-4}$ & 0.7$\pm$0.5 10$^{-4}$ \\
%    &   &   & Max. Intensity &  &     &   & 8.7 10$^{-4}$ & \\
 &  &  &  &  &   &   &  &   \\
  &   &   & Obs. Keck & 1.8$\pm$0.6 10$^{-4}$  &  2.2$\pm$0.7 10$^{-4}$ & 0.46$\pm$0.46 10$^{-4}$  & 4.7$\pm$2.3 10$^{-4}$  & 1$\pm$0.8 10$^{-4}$  \\
    &   &   & Max. Intensity &  &     &   & 13.6 10$^{-4}$  & \\
   & &  &   Corr. Extinction  & + 16\% & +16\% & +16\% &   +56\% & +56\%\\
 &  & &    Obs. NTT &  0.3 10$^{-4}$& 0.7 10$^{-4}$ & & 3.0 10$^{-4}$  & \\
&  & &   Meudon model & - & - & 0.1 10$^{-4}$   & 22 10$^{-4}$ &3.7 10$^{-9}$   \\
\hline
Br $\gamma$ & 2.16612 & 3.045 10$^5$ & 154582.7 &      &  &    &  & \\
 &  &  &  &  &   &   &  &   \\
% &  &  &  Obs. Keck &  13.4$\pm$0.1 10$^{-4}$ & 22.9$\pm$0.3 10$^{-4}$ &  8.0$\pm$0.7 10$^{-4}$  & 4.2$\pm$0.7 10$^{-4}$ &  2.3$\pm$0.7 10$^{-4}$\\
%&  &  &  Max. Intensity &  18.2 10$^{-4}$] & 28.2 10$^{-4}$ &  &  &  \\
 &  &  &  Obs. Keck &  15.4$\pm$0.1 10$^{-4}$ & 26.3$\pm$0.3 10$^{-4}$  &  9.2$\pm$0.8 10$^{-4}$   & 4.8$\pm$0.8 10$^{-4}$ &  2.6$\pm$0.8 10$^{-4}$  \\
&  &  &  Max. Intensity &  20.9 10$^{-4}$   & 32.4 10$^{-4}$   &  &  &  \\
  &   &   & Corr. Extinction  & +15\% & +15\%& +15\% & +15\%&+15\% \\
  &  &  &  Obs. NTT &  36.6 10$^{-4}$ & 39.4 10$^{-4}$ & & 11.7 10$^{-4}$  & \\
 & &  & Cloudy model &   4.7 10$^{-4}$    &        36 10$^{-4}$ &  - & - & - \\
%Brg
%HII : Median : 1.34e-03  -- std error: 1.46e-04 --25% error : 3.35e-04 -- Max  : 1.82e-03
%IF : Median : 2.29e-03  -- std error: 2.94e-04 --25% error : 5.72e-04 -- Max  : 2.82e-03
%atomic : Median : 8.06e-04  -- std error: 7.71e-05 --25% error : 2.02e-04 -- Max  : 1.12e-03
%DF : Median : 4.18e-04  -- std error: 7.13e-05 --25% error : 1.04e-04 -- Max  : 6.57e-04
%molecular : Median : 2.32e-04  -- std error: 6.85e-05 --25% error : 5.79e-05 -- Max  : 4.69e-04
%H2
%HII : Median : 1.62e-04  -- std error: 5.16e-05 --25% error : 4.05e-05 -- Max  : 4.03e-04
%IF : Median : 1.97e-04  -- std error: 5.97e-05 --25% error : 4.92e-05 -- Max  : 3.96e-04
%atomic : Median : 3.45e-05  -- std error: 4.33e-05 --25% error : 8.63e-06 -- Max  : 2.45e-04
%DF : Median : 2.99e-04  -- std error: 1.54e-04 --25% error : 7.48e-05 -- Max  : 8.71e-04
%molecular : Median : 7.39e-05  -- std error: 4.71e-05 --25% error : 1.85e-05 -- Max  : 2.27e-04
\hline
\hline
\end{tabular}
\end{table*}

\section{Comparison to model predictions}
\label{sec:model-predictions}

To guide our interpretation, in this section, we compare our extinction-corrected observations of the emission gas lines to the model predictions in the HII region, the IF, the atomic region, the DF, and the molecular zone. 
The model predictions are those we
 used to make the template spectra described in \citet{paper_ERS_PASP_2022} and to predict line intensities for the JWST Exposure Time Calculator. The gas line intensities have been computed  using (i) the Cloudy code for the ionized gas in the \HII~region and the ionization front \citep[][]{cloudy};
and (ii) the Meudon PDR code for the contribution from the atomic and molecular lines in the neutral PDR gas \citep[][]{lepetit}. The parameters and calculation requirements used for each model and region are described in detail in \citet{paper_ERS_PASP_2022}. %Only a very brief description is given here (see Sect.\ref{sec:cloudy-model} and \ref{sec:meudon-model}).
In the following subsections, only some computational results important for the comparison predictions/observations are given.

In this comparison with model predictions, our primary goal was to ensure that the models reproduce the observed intensity peak  of the ionised and neutral gas lines emission by Keck. This allowed us to adjust the required JWST integration time in order to get a high signal-to-noise ratio and avoid saturation problems, a challenging aspect for JWST observation of bright targets.
We emphasize that our model predictions were obtained by making simple assumptions and by separately estimating the ionized and neutral molecular gas lines. The radiative transfer and the thermal balance were not calculated continuously from the HII region to the molecular region. We have not attempted to reproduce the separation between the IF and the DF. 
%The Meudon code was used to model the H$_2$ line emission from the dense sub-structures. 

\subsection{Predicted Br$\gamma$ and [FeII] lines emission}
\label{sec:cloudy-model}

%For the ionized gas component, we adopt the same model parameters as in \citet{Shaw09} and \citet{Pellegrini09}.
%An illuminating star characterized by a Kurucz star model with an effective temperature $T_{eff}$~=~39,600~K was considered. 
%The total number of ionizing photons emitted by the star is set to \mbox{$Q_{\rm LyC}$\,$=$\,9.8$\cdot$10$^{48}$~photon\,\,s$^{-1}$}. 
%For the density, the initial electronic density is assumed to be $n^0_e$~=~3160~cm$^{-3}$, and a constant pressure was assumed. The total proton density at the IF increases until $n^$~=28,000~cm$^{-3}$.
The input parameters of \citet{Pellegrini09} and \citet{Shaw09} were adopted for the Cloudy model of the Orion Bar. This is an isobaric model with initial electron density of $n_e\sim~3500$ cm$^{-3}$ and temperature of $T_e\sim9000$ K. The electron density reaches a maximum of $n_e\sim$6600~cm$^{-3}$ before the IF and drops sharply at the IF. At the IF, the total proton density increases to $n$~=28,000~cm$^{-3}$ while the electronic temperature decreases to $T_e\sim$4000~K. 
The Br$\gamma$ 2.16 $\mu$m line emissivity is spatially extended in the HII region, following the electronic density profile,  and then decreases strongly when H$^+$ becomes H$^0$ at the IF. On the other hand, the [FeII] 1.64$\mu$m line emissivity peaks strongly at the IF. 
%Thermal pressure : 3500.*9000.=3e7 to 28000x4000=1.12000e+08  

In Table~\ref{tab_intensity},  we report the predicted intensities of the lines. % from the ionized, IF, atomic and H/H2 transition region and from the molecular region which starts at the C/CO transition. 
The [FeII] 1.64 $\mu$m and Br$\gamma$ 2.16 $\mu$m lines model predictions are in good agreement with the observations at the IF front (by a factor less than 1.4). On the other hand, the emission of these lines in the HII region are significantly lower in the model predictions. This difference applies for several other ionized gas lines predicted by Cloudy. % and observed with NTT/SOFI \citep{Walmsley00}.
The observations may overlap the IF where the observed intensities are bright.
%The considered HII region in the observations is close to the IF where the material density column increases and the measured HII lines intensity could be affected by the IF zone. %It should be noted that for the [FeII] line the  under-estimate by the model is much greater (a factor greater than 10).
%Considering that the physical conditions in the models fit the data, first constraint on the Fe abundance can be derived ?

\subsection{Predicted H$_2$ line emission}
\label{sec:meudon-model}

For the Meudon code, we considered an isobaric model with a thermal pressure \mbox{$P=4\cdot 10^8$\,K\,cm$^{-3}$}. This model exhibits gas physical conditions expected near the dissociation front of the Orion Bar \citep[e.g.,][]{Goico16,Joblin18}. We fixed the radiation field impinging on the PDR so that, at the edge of the PDR, $G_0\,=\,2\cdot10^4$ in Mathis units.  %We adopted the extinction curve of HD~38087 of \citet{Fitzpatrick1990} and $R_V\,=\,5.62$  close to the value determined for Orion Bar of 5.5 \citep{Marconi98}+Cardelli. 
We adopted the extinction curve of HD~38087 of \citet{Fitzpatrick1990} and $R_V\,=\,5.62$ which is
close to the value determined for Orion Bar of 5.5 \citep{Marconi98}. 
The model includes an exact radiative transfer calculation for the UV pumping lines originating from the first 30 levels of H$_2$, while the other lines are treated using the FGK approximation. %(Federman et al. 1979). 
This allows to account for mutual shielding effects between overlapping H$_2$ lines and can significantly affect the position of the H/H$_2$ transition.
%We carried out an exact radiative transfer calculation of the FUV radiation field \citep{lepetit}.A complete transfer with exact calculation of the mutual screening between the UV pumping lines of H$_2$ was done since it has a significant effect on the position of the H/H$_2$ transition and influences the intensities of the ro-vibrational lines of the radiative cascade. This  calculation increases in fact the 1-0 S(1) line intensity by a factor of about $\sim$3. 
The H$_2$ 1-0 S(1) line emissivity peaks near the H/H$_2$ transition layer where the gas density is of the order of few 10$^5$ cm$^{-3}$. As in \cite{Joblin18}, the model does not reproduce the observed width of the atomic region, i.e. the distance between the position of the IF and the DF.  
It is significantly smaller than in the observation. 
%The H/H$_2$ transition is at $A_V=0.3-0.4$ from the PDR edge corresponding to $\sim$0.001~pc or $\sim$1.5''. %In the PDR model, the PDR begins almost where the emission of H$_2$ is observed. %Each observed bright substructure in H$_2$ corresponds to a PDR model. 
%This is much smaller than in the observation. 
We note that as suggested by \cite{Allers05}, a PDR model with a lower gas pressure in the atomic zone (i.e., a lower gas density) and a reduction of the FUV dust cross section will better match the observed spatial position of the H$_2$ emission peak.
In the \cite{Schirmer2022} dust model with nano-grain depletion (i.e., FUV extinction reduction), the visual extinction integrated from the PDR edge to the observed H$_2$ emission peak position at 12-12.5$''$ (or 0.024-0.025~pc) is of $A_V\sim 1$. The required model medium density in the atomic zone is of 5 10$^4$ cm$^{-3}$, close to the inter-clump density.  %This extinction corresponds to that of a layer with a gas density of 5 10$^4$ cm$^{-3}$, a width of $\sim$0.02~pc, and a $A_V /N_H=3.5 \times 10^{-22}$ mag cm$^{-2}$.   %5e4*0.02*3.09e18*3.5e-22 =1
%The model density in the atomic zone should thus be close the inter-clump density.
%The model pressure in the atomic zone could thus be close the inter-clump pressure.

We calculated the  cumulative line intensities from the atomic and H$^0$/H$_2$ transition region (0$<A_V<$0.4 and 0.4$<A_V<$2.5) and from the molecular region (which starts at the C/CO transition, $2.5<A_V<8.5$). %We set an upper limit of $A_V$=8.5 to the molecular region to eliminate emission caused by the interstellar radiation field on the back side.
The optically thin line surface brightness are enhanced by a geometrical factor of  1/cos($\theta$)=2 relative to the face-on surface brightness.
The viewing angle $\theta$ between the line-of-sight and the normal to the PDR is taken equal to $\sim$60$^{\circ}$ which gives an approximation of a nearly edge-on PDR and is the maximum inclination that can be used to derive line intensities in the 1D PDR Meudon code. The uncertainty on this angle could lead to an additional scaling factor on the line intensity. %This PDR model provides a good agreement with the observed line intensities of the high-J CO emission produced in the warm dense layers before the C$^+$/C/CO transition \citep{Joblin18}. %A global scaling factor of 1.3 was found to fit the observed lines intensity. %The typical predicted thickness of the high-J CO lines emission is of a few 10$^{-3}$\,pc or $\sim$1''.  

%To reproduce the nearly edge-on geometry of the Bar, we adopt a geometry in which the PDR is observed with a viewing angle $\theta$ between the line-of-sight and the normal to the PDR equal to 60$\sim$60$^{\circ}$. This angle is being defined with 0$^{\circ}$ being face-on and 90$^{\circ}$ edge-on. The value of 60$^{\circ}$ gives an approximation of a nearly edge-on PDR and is the maximum inclination that can be used to derive line intensities in the 1D PDR Meudon code.
%Note: \citep{Joblin18} considered this geometrical factor and allowed for a free global scaling factor on the model intensities when fitting the model. a global scaling factor of 1.3 was found. 

The H$_2$ 2.12 $\mu$m line intensity as predicted by the model in the DF is in agreement with the measured peak intensity at the DF by a factor of 1.6 (too high for the model). This could partly be due to a geometrical factor but also to the fact that the isobaric model considered here with a high pressure of $P=4~10^8$ K cm$^{-3}$ gives a too high density and over-estimates the H$_2$ 1-0 S(1) line (see section \ref{sec-model-pressure}). 
%As discussed below from TEXES observations, a lower pressure could be required to account the H$_2$ lines intensity ratio.
The H$_2$ 1-0 S(1) line emissivity width predicted by the model is of $\sim$0.0006~pc or $\sim$0.3''. This is lower that the observed thickness of the sharpest H$_2$ filaments (see Fig.~\ref{fig:Keck-cut}). The observed width of the bright H$_2$ ridge (about 1$''$ to 2$''$) may be increased by geometrical effects or a lower thermal pressure. %Geometrical effect=Considering an angle of 7deg and a length a long the line of sight of 0.02 pc = sin(10.*!pi/180.)*0.02 = 0.0024 pc = 1.2'' 
%Javier : Why dont you directly run a isobaric model a give the pressure for which the line intensity is reproduced?
%Emilie : because we have to fit other H2 lines - 1 line is not sufficient
%Anyways, a factor 1.6 does not seem to high for me. Joblins model fits the high-J CO lines, which have high critical densities. This is a nice conclusion: that H2* and high-J CO lines seem to require different pressures, meaning that is not exactly the same gas component????

In the atomic zone, the model predicts a much lower intensity by a factor of $\sim$10 with respect to the DF because the H$_2$ density is much lower. This is in agreement with the data but the large uncertainties on the measured intensity make a quantitative comparison difficult. 
%there is also a relative good agreement between the H$_2$ line observation and prediction by a factor of.
In the molecular zone, the predicted intensities are extremely low by a factor $>10^4$ with respect to the DF. The measured intensity decreases much less. It is comparable to the intensity measured in the HII region divided by a factor of 2. 
%Thus, the observed H$_2$ emission in the molecular zone must come from foreground emission, from the face of the Bar or from other foreground layer.
The observed H$_2$ emission in the molecular zone must come from the flattened region behind the main ionization front on the 
far side of the Orion Bar.
% \ref{sec:spatial-distribution-ionised-gas}.

\subsection{Thermal pressure and dynamic effects}
\label{sec-model-pressure}

Considering the TEXES H$_2$ 0-0 S(1) line observation with an angular resolution of 2$''$ \citep{Allers05} and the 1-0 S(1) line Keck observation convolved at the same resolution and corrected for extinction, the 1-0 S(1) / 0-0 S(1)  line ratio is estimated to be about $\sim $1$\pm 0.4$ at the DF (position A in Allers et al. 2005). The ratio predicted by the model is about 2.5 times higher. 
%For a model with a lower pressure, i.e. a lower density, this ratio will decrease and will be closer to the data. For instance, in an isochoric PDR model with $n=10^5$ cm$^{-3}$ the 1-0 S(1)/0-0 S(1)  line ratio is predicted to be about 0.23 which is 10 times lower than the high pressure model.
A pressure of $4~10^8$ K~cm$^{-3}$ is  probably too high to reproduce this ratio and more detailed modeling would be needed. 
%This suggests that the pressure in the H$_2$ emission zone must be lower than the one derived for the high-J CO lines. 
%Therefore it seems that a single pressure is not sufficient to describe the evolution of gas density and temperature across the hot/warm irradiated interface. 
The previous H$_2$ study by \cite{Allers05} show in fact that an isobaric model with  $P=8~10^7$ K cm$^{-3}$ (with a pressure 5 times lower that the one we assumed) matches the observed intensities of the ground-state rotational lines, the H$_2$ 1-0 S(1) and 2-1 S(1) lines. %The distance between the H$_2$ line emission and the IF is moreover well reproduced by this isobaric model assuming a reduction in the FUV attenuation cross section.  
%it is apparent that the standard dust-related parameters used in their PDR models do not allow for a reasonable match to their observations. In order to reproduce the observed separation between the ionization front and the H2 emission peak, their model requires a reduction in the FUV attenuation cross section by a factor of 3 relative to a priori estimates. This is in agreement with Schirmer et al. reduction de la bosse a 4.2 microns-1, courbe plus plate au-dela de la bosse (and also requires an en- hanced heating rate in the atomic region of the PDR, corresponding to a factor of 3 increase in the photoelectric heating rate.) 

To determine the pressure and density variations at the PDR edge, 
future detailed spatial studies of both the H$_2$  pure rotational and rovibrational lines are required.
Excitation diagram from the numerous H$_2$ lines observable by JWST MIRI and NIRSpec will permit to trace the warm and hot (UV-pumped) excitation temperatures, local gas density and pressure gradients at small spatial scales. It will be possible to probe the physical conditions in most of the dense sub-structures detected with Keck or ALMA as well as the very thin surface layers that are sufficiently heated to photo-evaporate from the PDR. On the other hand, one could marginally spatially resolve the temperature gradient inside the over-dense substructures themselves. For a high pressure PDR, the  H$_2$ emission thickness is predicted to be $<$1'' and spatial shift between the H$_2$ rotational line emission peaks is comparable or smaller to the JWST diffraction limit which is about 0.2'' to 1'' in the 5-28 $\mu$m range where are the pure rotational lines. 
%The thickness of the H$_2$ emission region as predicted by the model is about 0.5''. 

These constraints on the physical conditions may allow to better understand the dynamical effects in PDRs, e.g. compression waves, photo-evaporative flows, and IF and DF instabilities.
%We must underlines that dynamic and non-equilibrium effects are important aspects to consider for data interpretation
Recently, \cite{Maillard2021} modelled the dynamical effects of the radiative stellar feedback on the H-to-H$_2$ transition. These authors built a semi-analytical model of the H/H$_2$ transition in a 1D plane-parallel PDR illuminated by FUV radiation with advancing IF resulting from the photoevaporation mechanism. For the high-excitation PDRs such as the Orion Bar, moderate  effects are however predicted. The H$^0$/H$_2$ transition will  be slightly closer to the IF and sharper. Much stronger effects of the advection dynamics with merged, or close to merged, IF and DF are predicted for low excitation PDRs illuminated by late O stars such as the Horsehead nebula.

%Confirmation of the presence of small dense structures of typical thickness of a few 1e-3 pc and at high thermal pressures (Pth=10$^{18}$  K  cm$^{-3}$)  as  suggested  the  analysis  of  high-J  CO  lines(by  Joblin  et  al.  2018)

\section{Summary}
\label{sec:conclusions}

In this section, a short summary and prospects for JWST are given. 

\begin{enumerate}

\item We obtained with Keck/NIRC2 using AO the most complete and detailed maps of the complex UV-irradiated region of the Orion Bar where the conversion from ionized to atomic to molecular gas occurs.
We mapped the vibrationally excited line of H$_2$ at 2.12 $\mu$m, tracing the dissociation front, and the [FeII] line at 1.64 $\mu$m and the Br$\gamma$ line at 2.16 $\mu$m, tracing the ionization front.
This allows for the first time to spatially resolve the dissociation front with a resolution of 0.1$''$ (40 AU) and obtain a very precise determination of the offset between the peak of the ionization and dissociation front of 12 and 12.5$''$ (0.024-0.025 pc).
%The ionization front is clearly seen in the Br? and [FeII] line emission maps.

\item  The H$^0$/H$_2$ dissociation front appears highly structured with several ridges and extremely sharp filaments with a width of $1-2''$ (0.002-0.004~pc or 400-800~AU). Ridges and filaments run parallel to the dissociation front but a succession of bright sub-structures in H$_2$ is also observed from the edge of the DF towards the molecular region. The multiple H$_2$ emission peaks along and across the Bar may be associated with a multitude of small highly irradiated and dense PDRs.
%Series of ridges that follow along the interfaces may be associated with a multitude of small dense highly irradiated PDRs. 

\item The comparison with ALMA data of the HCO$^+$ J=4-3 line emission, a good indicator of dense gas, show a remarkable similar spatial distribution between the H$_2$ and HCO$^+$ emission. This suggests that they both come from high densities but also indicates  their strong chemical link and  that the H/H$_2$ and C$^+$/C/CO transition zones are very close.
 
\item We spatially resolved additional compact sources corresponding to over-dense irradiated sub-structures such as proplyds near the ionization front, and perhaps collapsing protostars near the dissociation front. 

\item We compared our extinction-corrected observations of the emission gas lines to the model predictions using the Cloudy code for the ionized gas and the Meudon PDR code for the neutral PDR gas. The models reproduce well (by a factor less than 2)  the observed intensity peak of the gas lines emission. The predicted vibrationaly excited H$_2$ come from the H$^0$/H$_2$ transition layer where the gas density is of the order of few 10$^5$ cm$^{-3}$. % (for isobaric models P=1e8 K/cm-3)
%The [FeII] 1.64 ?m and Br? 2.16 ?m lines model predictions are in good agreement with the observations at the IF (by a factor less than 1.4). and the H2 2.12 ?m line intensity as predicted by the model in the DF is in agreement with the measured peak intensity at the DF (by a factor less than 1.6)

\item This work confirms without ambiguity that one of the densest portions of the Bar lie along the dissociation front and  that the Orion Bar edge is composed of small, dense sub-structures at high thermal pressure with P about 10$^8$~K~cm$^{-3}$ immersed in a more diffuse environment.
% as suggested by the modeling of the warm dense molecular gas tracers  observed with Herschel (e.g., high-J CO) and ALMA (e.g., HCO$^+$).
%The UV shielding produced by the ridge of high-density substructures may  contribute to protecting the molecular cloud from photodestruction.
%However, a single pressure is probably not sufficient to describe the evolution of gas density and temperature across the hot/warm irradiated interface. 
Further studies are required to accurately determine the gas physical conditions (pressure and density gradients) and dynamics of these over-dense irradiated sub-structures and better understand their
physical origin and future.

\end{enumerate}

The JWST ERS proposal on the Orion Bar PDR \citep{paper_ERS_PASP_2022} will very soon give access to multiple spectroscopic images with a similar spatial resolution to the Keck/NIRC2 observations but a signal-to-noise gain that will increased contrast and allow a sharper detection of all less bright substructures.
JWST with IFU spectroscopy will  give insight into the local gas physical conditions (temperature, density and pressure) and chemical composition of the warm very structured irradiated medium. 
It will be possible to probe the physical conditions in most of the dense sub-structures detected with Keck or ALMA as well as the very thin surface layers that are sufficiently heated to photo-evaporate from the Orion Bar edge.
Furthermore, JWST will spatially resolve at near and mid-IR wavelengths dust emission and scattering simultaneously with the gas lines. 
Variation of the dust properties (e.g., size distribution, extinction) as function of PDR depth will be constrained  and will be taken into account in the UV shielding, thermal balance and H$_2$ formation process.

{\it Acknowledgements.}
%{\bf We thank our referee, Robert OÕDell , for the care and quality of his comments and suggestions that improved the clarity of this paper. }
The data presented herein were obtained at the W. M. Keck Observatory, which is operated as a scientific partnership among the California Institute of Technology, the University of California and the National Aeronautics and Space Administration. The Observatory was made possible by the generous financial support of the W. M. Keck Foundation.

The authors wish to recognize and acknowledge the very significant cultural role and reverence that the summit of Maunakea has always had within the indigenous Hawaiian community.  We are most fortunate to have the opportunity to conduct observations from this mountain.

This research has made use of the Keck Observatory Archive (KOA), which is operated by the W. M. Keck Observatory and the NASA Exoplanet Science Institute (NExScI), under contract with the National Aeronautics and Space Administration.
Javier R. Goicoechea thanks the Spanish MCIYU for funding support under grant PID2019-106110GB-I00.

\bibliographystyle{aa}
\bibliography{biblio.bib}

%\bibliographystyle{aa}
%\bibliographystyle{aasjournal}
%\bibliography{biblio.bib}

\end{document}